\documentclass[]{pasj02} 
\usepackage[switch,mathlines]{lineno}
\usepackage{natbib} 
\usepackage{lscape}
\usepackage{url}
\usepackage{rotating}
\usepackage{graphicx} 
\usepackage{ulem}
\usepackage{xcolor}
\jyear{2025}
\Received{2025/12/18}
\Accepted{2026/03/19}

\newcommand{\oi}{[O {\sc i}]}
\newcommand{\oiii}{[O {\sc iii}]}
\newcommand{\ci}{[C {\sc i}]}
\newcommand{\cii}{[C {\sc ii}]}
\newcommand{\nii}{[N {\sc ii}]}
\newcommand{\niii}{[N {\sc iii}]}

\begin{document} 

\title{
ATT12: The Antarctic 12-m Terahertz Telescope for Studies of Dusty Galaxies. I. Instrument Sensitivity and Science Forecasts 
}

\author{
 Koki Wakasugi,\altaffilmark{1}
 Takuya Hashimoto,\altaffilmark{1,2}\orcid{0000-0002-0898-4038}\altemailmark \email{hashimoto.takuya.ga@u.tsukuba.ac.jp}
 Nario Kuno,\altaffilmark{1,2}\orcid{0000-0002-1234-8229}
 Yu Nagai,\altaffilmark{1}
 Naomasa Nakai,\altaffilmark{3}\orcid{0000-0002-5461-6359}
 Ken Mawatari,\altaffilmark{1,2,4,5}\orcid{0000-0003-4985-0201}
 Masumichi Seta,\altaffilmark{3}\orcid{0000-0002-9450-6779}
 Shun Ishii,\altaffilmark{6,8}\orcid{0000-0001-8337-4961}
 %
 Shunsuke Honda,\altaffilmark{1,2}\orcid{0000-0002-0403-3729} 
 Mana Ito,\altaffilmark{1,7}\orcid{0009-0002-8249-9353}
 Hiroshi Matsuo,\altaffilmark{6,8}\orcid{0000-0003-3278-2484}
 Makoto Nagai,\altaffilmark{6}\orcid{0000-0002-1705-9610}
 Yuri Nishimura\altaffilmark{1,2,12}\orcid{0000-0003-0563-067X}
 Dragan Salak,\altaffilmark{9,10}\orcid{0000-0002-3848-1757}
 Kazuo Sorai,\altaffilmark{10,11}\orcid{0000-0003-1420-4293}
 and 
 Hidenobu Yajima\altaffilmark{1,2,7}\orcid{0000-0002-1319-3433}
 }
\altaffiltext{1}{
Division of Physics, Faculty of Pure and Applied Sciences, University of Tsukuba, Tsukuba, Ibaraki 305-8571, Japan
}
\altaffiltext{2}{
Tomonaga Center for the History of the Universe (TCHoU), Faculty of Pure and Applied Sciences, University of Tsukuba, Tsukuba, Ibaraki 305-8571, Japan
}
\altaffiltext{3}{
Department of Physics \& Astronomy, School of Science, Kwansei Gakuin University, 1 Gakuen Uegahara,, Sanda, Hyogo 669-1337, Japan
}
\altaffiltext{4}{
Waseda Research Institute for Science and Engineering, Faculty of Science and Engineering, Waseda University, 3-4-1 Okubo, Shinjuku, Tokyo 169-8555, Japan
}
\altaffiltext{5}{
Department of Pure and Applied Physics, School of Advanced Science and Engineering, Faculty of Science and Engineering, Waseda University, 3-4-1 Okubo, Shinjuku, Tokyo 169-8555, Japan
}
\altaffiltext{6}{
National Astronomical Observatory of Japan, 2-21-1 Osawa, Mitaka, Tokyo 181-8588, Japan
}
\altaffiltext{7}{
Center for Computational Sciences, University of Tsukuba, Ten-nodai, 1-1-1 Tsukuba, Ibaraki 305-8577, Japan
}
\altaffiltext{8}{
Graduate University for Advanced Studies (SOKENDAI), 2-21-1 Osawa, Mitaka, Tokyo 181-8588, Japan
}
\altaffiltext{9}{
Institute for the Advancement of Higher Education, Hokkaido University, Kita 17 Nishi 8, Kita-ku, Sapporo, Hokkaido 060-0817, Japan
}
\altaffiltext{10}{
Department of Cosmosciences, Graduate School of Science, Hokkaido University, Kita 10 Nishi 8, Kita-ku, Sapporo, Hokkaido 060-0810, Japan
}
\altaffiltext{11}{
Department of Physics, Faculty of Science, Hokkaido University, Kita 10, Nishi 8, Kita-ku, Sapporo, Hokkaido 060-0810, Japan
}
\altaffiltext{12}{
Tsukuba Institute for Advanced Research (TIAR), University of Tsukuba, 1-1-1 Tennodai, Tsukuba, Ibaraki, 305-8577, Japan
}



\maketitle

\begin{abstract}
We present a feasibility study of the Antarctic 12-m Terahertz Telescope (ATT12), a next-generation facility to be constructed at New Dome Fuji in Antarctica, designed to open up the far-infrared and terahertz windows for extragalactic astronomy.
While ATT12 will enable a wide range of Galactic and extragalactic science, this paper focuses on its potential for studies of dusty star-forming galaxies (DSFGs) across cosmic time. 
Using realistic atmospheric transmission models and the planned instrumental specifications of heterodyne spectrometers and wide-field multi-color continuum cameras, we assess the expected sensitivity and scientific capabilities of ATT12.
We show that spectroscopic observations will enable detections of \cii~158\,$\mu$m from galaxies with $\log(L_{\rm IR}/L_\odot)\gtrsim12$ out to $z\sim7$, while \oiii~88\,$\mu$m will remain observable for HyLIRG-class systems ($\log L_{\rm IR}/L_\odot\gtrsim13$) up to $z\sim10$.
Line ratios including \oiii~52/88\,$\mu$m, \nii~122/205\,$\mu$m, and \oiii/\niii\ will provide unique diagnostics of electron density and O/N abundance at $z\sim4$--8.
Wide-field continuum surveys with the continuum cameras (KIDS-1/2; 300--850~GHz) will reach confusion-limited depths of $\sim$1--2~mJy over $\sim$10,000~deg$^2$, detecting of order $10^{6}$--$10^{7}$ dusty star-forming galaxies with $L_{\rm IR}\gtrsim10^{12}L_\odot$ at $z\lesssim5$ and $\lesssim10^{3}$--$10^{4}$ HyLIRGs with $L_{\rm IR}\gtrsim10^{13}L_\odot$ up to $z\sim7$ or even higher.
Higher-frequency MKID cameras (KIDS-3/4; $>$850~GHz) are designed for targeted follow-up observations and to extend coverage toward the THz regime.
Taken together, ATT12 will provide the first statistically representative samples of dusty galaxies across cosmic time and, through synergy with ALMA, JWST, and the proposed FIR Probe PRIMA, will establish a multi-wavelength framework in which ATT12 discovers large samples through wide-area surveys, ALMA provides high-resolution follow-up of gas and ISM structure, JWST probes stellar populations and metallicity in the rest-frame optical/near-infrared, and PRIMA delivers ultra-sensitive far-infrared spectroscopy from space. 
\end{abstract}


\section{Introduction}
\label{sec:intro}

\subsection{Scientific Background}
\label{subsec:intro1}

Understanding the formation and evolution of galaxies across cosmic time is one of the central goals of modern astrophysics.
In particular, the buildup of dust and metals and their impact on star formation remain key open questions.
The cosmic star formation rate density (SFRD) peaks around $z\sim2$ (``cosmic noon''), when galaxies experienced intense star formation and rapid enrichment \citep[e.g.,][]{Madau2014}, coincident with the peak of active galactic nucleus (AGN) activity and rapid supermassive black hole growth \citep[e.g.,][]{Hopkins2006,Ueda2014}.

While optical/near-infrared surveys map the unobscured component, a substantial fraction of star formation is hidden by dust, especially in the most luminous infrared galaxies \citep[e.g.,][]{Casey2014}.
Constraining the role of dusty star-forming galaxies (DSFGs) is therefore essential.
An example of distant DSFGs is the population of submillimeter galaxies (SMGs; e.g., \citealt{Smail1997, Hughes1998, Blain2002,Casey2014}), massive, dust-enshrouded systems with SFRs of $\sim$100–1000~$M_{\odot}$~yr$^{-1}$ (e.g., \citealt{Weiss2013,Strandet2016,Reuter2020}).
Heavy obscuration often renders their activity invisible at optical/UV wavelengths, though recent \textit{JWST} observations are beginning to provide new insights \citep{Arribas2024,Jones2024,AlvarezMarquez2023,CrespoGomez2023,Colina2023,Hodge2025}.

Far-infrared (FIR) and submillimeter observations provide a direct probe of dust-obscured star formation and the interstellar medium (ISM). 
Early constraints on the FIR luminosity function (LF) in the local universe were obtained from IRAS all-sky surveys \citep{Soifer1987,Saunders1990}.
Wide-area surveys with the \textit{Herschel Space Observatory (Herschel)} later revealed thousands of DSFGs and provided the first measurements of the FIR LF at higher redshifts ($z\gtrsim1$) \citep{Eales2010,Oliver2012}, although they were still affected by source confusion at longer wavelengths.
Complementary JCMT/SCUBA(–2) surveys played a crucial role in identifying high-$z$ DSFGs and constraining the bright end of the IR LF \citep[e.g.,][]{Smail1997,Coppin2006,Geach2017,Michalowski2017}.
SPT extended the census to millimeter wavelengths \citep{Carlstrom2011,Reuter2020}, but with coverage largely on the Rayleigh–Jeans side of the SED.
Recent ALMA programs advanced IR~LF and dust-obscured SFRD measurements across cosmic time: ALCS derived IR~LFs to $z\sim8$ \citep{Fujimoto2024}; \citet{Zavala2021} and \citet{Barrufet2023} constrained the evolution of dust-obscured star formation and reported the first IR~LF at $z\sim7$; multiwavelength work further refined IR~LF/SFRD \citep[e.g.,][]{Gruppioni2020}. 
Yet ALMA’s small field of view limits surveyable sky, leaving the global abundance and contribution of DSFGs—particularly at $z>4$—to the SFRD during reionization still uncertain \citep[e.g.,][]{RowanRobinson2016,Dudzeviciute2020}.

These observational efforts laid the groundwork for dusty star formation, but also highlight the need for deeper insight into the ISM.
FIR/submillimeter spectroscopy with ISO, \textit{Herschel}, \textit{AKARI}, ALMA, and SPT is far less affected by extinction and probes the internal structure and physical conditions of SMGs \citep[e.g.,][]{Hodge2016,Hodge2019,Tadaki2018,Nagao2012}.
Thus, FIR/submillimeter observations are indispensable for a comprehensive view of galaxy evolution.

One of the key advantages of the FIR--submillimeter window is access to bright fine-structure lines that diagnose ionizing sources (star formation vs.\ AGN) and ISM conditions (temperature, density, ionization state).
In PDRs, strong lines such as \ci~371, 605\,$\mu$m, \cii~158\,$\mu$m and \oi~63, 145\,$\mu$m are emitted; {\sc Hii} regions produce \oiii~52, 88\,$\mu$m, \nii~122, 205\,$\mu$m, and \niii~57\,$\mu$m.
The \cii~158\,$\mu$m and \oi~63\,$\mu$m transitions are primary PDR coolants \citep{Hollenbach1999,Wolfire2022}, and \oiii~88\,$\mu$m traces ionizing-photon production \citep{Ferkinhoff2010,Inoue2014}.
Local calibrations link these lines to SFRs \citep[e.g.,][]{DeLooze2014}, and the \cii--SFR relation extends to $z\sim6$ \citep{Schaerer2019}.
Density-sensitive \oiii~52/88 and \nii~122/205 constrain electron density, while \niii~57/(\oiii~52+\oiii~88) probes N/O \citep[e.g.,][]{Nagao2012,Kewley2019,Pereira-Santaella2017,Spinoglio2022}.

Despite this diagnostic power, current ALMA samples remain small due to limited survey speed, and the global picture of ISM evolution in DSFGs across cosmic time is still incomplete.
In particular, understanding ISM evolution in the most luminous dusty starbursts is essential for revealing their origin and fate. These extreme systems—likely progenitors of today’s massive quiescent/elliptical galaxies—offer laboratories for studying how rapid gas consumption and feedback drive the transition from starburst to passive phases \citep[e.g.,][]{Toft2014,Simpson2014,Michalowski2010,Valentino2020}.

A new facility surveying large sky areas in the FIR--THz regime is required.
The Antarctic 12-m Terahertz Telescope (ATT12) is designed to combine wide-area coverage, access to unique Antarctic THz windows, and sensitivity to detect statistically significant DSFG samples to $z>6$.
In synergy with ALMA, JWST, and the proposed \textit{FIR Probe PRIMA}, ATT12 will connect detailed, small-scale studies with a global census of dusty galaxies.
We present the expected performance of ATT12 and its scientific potential for dusty-galaxy studies.

\subsection{The Antarctic 12-m Terahertz Telescope (ATT12)}
\label{subsec:intro2}
Building on this motivation, we outline ATT12’s site, instrumentation, and survey capabilities.

\begin{figure}[!t]
\includegraphics[width=0.45\textwidth]{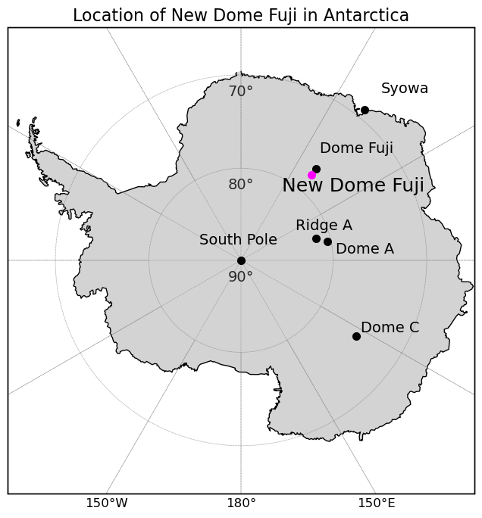}
\caption{Location of New Dome Fuji in Antarctica (magenta circle) together with other astronomical sites (black circles). Adapted from \citet{Ishii2010} and \citet{Saunders2009}. }
\label{fig:locations}
\end{figure}

The Antarctic Astronomy Consortium, led by the University of Tsukuba, plans a next-generation 12-m THz telescope (PIs: N. Nakai and N. Kuno) at New Dome Fuji, $\sim$50~km south of Dome Fuji station.
Figure~\ref{fig:locations} shows the site relative to the South Pole, Dome~A/C, and Ridge~A \citep{Ishii2010,Saunders2009}.
Dome Fuji station lies at 77$^\circ$19$'$S, 39$^\circ$42$'$E, altitude 3810~m; New Dome Fuji is comparably high ($\sim$3800~m) but farther inland, likely providing superior atmospheric conditions.

The high, dry interior of Antarctica is among the best sites for submillimeter/THz astronomy \citep[e.g.,][]{Peterson2003,Saunders2009,Burton2015,Shi2016,Ishii2010,Matsuo2019}, enabling high transparency at ALMA Bands~9/10 and into THz windows inaccessible from Atacama in typical conditions.

ATT12 will host
(i) a heterodyne spectroscopic system and
(ii) a wide-field imaging system with multi-band continuum cameras. Microwave Kinetic Inductance Detector (MKID) is regarded as a promising candidate for implementation as the continuum cameras.
As summarized in table~\ref{tab:heterodyne_specs}, the heterodyne system spans 275--2000~GHz ($\lambda \approx 150$~$\mu$m--1.1~mm) with multiple beams ($>100$~beams), and is designed to accommodate a dedicated VLBI front-end.
The optional VLBI front-end would further enable participation in global submillimeter VLBI arrays in the future, expanding ATT12’s science reach toward horizon-scale imaging and high-frequency astrometry.
We focus here on the multi-beam spectroscopic capabilities.

As shown in table~\ref{tab:kids_specs}, the KIDS camera operates from 300 to 2000~GHz ($\lambda \approx 150$~$\mu$m--1.5~mm), complementing the heterodyne receivers with large-area surveys.

The lower frequency limit was set to fully exploit Antarctic advantages, while the upper limit is set by available windows (optical depth $\tau<3$ at $40^\circ$ elevation), surface accuracy, and pointing requirements.

Because surface imperfections and illumination patterns reduce the effective collecting area, the effective (rather than geometric) aperture must be considered.
The aperture efficiency is
\begin{equation}
\eta_{\rm A}=\eta_0 \exp\!\left[-\left(\frac{4\pi\epsilon}{\lambda}\right)^2\right],
\end{equation}
with $\eta_0$ set by illumination, $\epsilon$ the rms surface error, and $\lambda$ the wavelength \citep{Ruze1966,Wilson2009}.
We adopt $\eta_0=0.7$, representative of a Cassegrain telescope without subreflector blockage and with a $\sim$1~deg$^{2}$ field of view \citep{Nagai_Imada2024}.
To observe up to 2~THz ($\eta_{\rm A}\gtrsim0.15$), we aim for $\epsilon \approx 15$~\micron\ for the primary reflector.

We target $0\farcs5$ pointing/tracking accuracy.
The HPBW is
\begin{equation}
\theta_{\mathrm{HPBW}}=K\frac{\lambda}{D},
\end{equation}
with $K=1.2$ (Gaussian beam), $\lambda$ the observing wavelength, and $D$ the antenna diameter.
At 850, 1000, and 2000~GHz the beams are $7\farcs3$, $6\farcs2$, and $3\farcs1$ (table~\ref{tab:kids_specs}).

KIDS surveys will pre-select candidates via photometric redshifts \\bf{\citep[e.g.,][]{Hughes2002, Aretxaga2003,Ivison2016,Casey2020}, which will be presented in a forthcoming paper.
The optical design targets $>$1~deg$^2$ FoV.
With KIDS-1/2, surveys will cover $\sim$10,000~deg$^2$ south of $\delta=0^\circ$, excluding the Galactic plane.
Reaching the confusion limit is expected to require about two years for KIDS-1 (300/400/500~GHz) and $\sim$eight years for KIDS-2 (650/850~GHz)\footnote{Bands are arranged separately on the focal plane; mapping-time estimates assume a representative observing efficiency.} (table~\ref{tab:kids_specs}).
We adopt confusion values from \citet{Blain2002}; the maximum occurs at 500~GHz, corresponding to $\approx$1.7~mJy (5$\sigma$).
Promising candidates will be followed up at higher frequencies with KIDS-3/4.

This exceeds current surveys such as SPT ($\sim$2500~deg$^2$ at 95/150/220~GHz; \citealt{Carlstrom2011,Reuter2020}), optimized for SZ and bright DSFG discovery (few mJy at 150~GHz).

By contrast, \textit{Herschel} executed multi-tier extragalactic surveys spanning several hundred deg$^2$ with PACS/SPIRE, from wide/shallow (H-ATLAS, $\sim$600~deg$^2$; \citealt{Eales2010}) and HerMES ($\sim$380~deg$^2$; \citealt{Oliver2012}) to deeper programs (HeLMS, HerS, PEP, GOODS-Herschel; \citealt{Asboth2016,Viero2014,Lutz2011,Elbaz2011}).
Overall coverage reached $\sim$700–800~deg$^2$ in the 100–500~$\mu$m range, with confusion-limited depths of a few mJy in the deepest fields.
However, \textit{Herschel} lacked all-sky coverage and wavelengths $>$500~$\mu$m, leaving discovery space for ATT12, which uniquely combines wide area with THz access (table~\ref{tab:survey_comparison}).

\begin{table*}[t]
\tbl{Specifications of the planned heterodyne receivers for ATT12.}{%
\centering
\begin{tabular}{lcc}
\hline
Spectroscopic receiver & Tuning range [GHz]\footnotemark[$*$] & Comment \\
\hline\hline
Receiver V\footnotemark[$\dag$] & 82--116, 210--280, 275--375 & VLBI (single beam) \\
Receiver 1 & 275--370, 385--540 & Multi-beam \\
Receiver 2 & 575--735, 775--965 & Multi-beam \\
Receiver 3 & 1000--1060, 1250--1380, 1450--1550 & Multi-beam \\
Receiver 4 & 1935--2000 & Multi-beam \\
\hline
\end{tabular}
}
\begin{tabnote}
  \begin{minipage}[t]{0.7\textwidth}
  \raggedright
  \footnotemark[$*$] Tuning ranges correspond to windows avoiding strong atmospheric absorption.\\
  \footnotemark[$\dag$] A dedicated VLBI receiver is also being considered; the frontend would be shared and the backend would be optimized for VLBI. The detailed specifications are still under development.
  \end{minipage}
\end{tabnote}
\label{tab:heterodyne_specs}
\end{table*}

\begin{table*}[t]
\tbl{Specifications of the KIDS bands (MKID-based continuum cameras).}{%
\centering
\begin{tabular}{lcccccc}
\hline
KIDS\footnotemark[$*$] & Central freq. [GHz] & HPBW [$^{\prime\prime}$] & Bandwidth [GHz] & $\Delta S$/1h (5$\sigma$) [mJy]\footnotemark[$\dag$]  & Confusion (5$\sigma$) [mJy] & Note \\
\hline\hline
KIDS-1 & 300  & 20.6 & 35 & 0.88  & 1.4              & Southern all-sky survey \\
       & 400  & 15.5 & 40 & 1.16  & 1.5              & Southern all-sky survey \\
       & 500  & 12.4 & 50 & 1.94  & 1.7              & Southern all-sky survey \\\hline
KIDS-2 & 650  &  9.5 & 65 & 1.89  & 1.3     & Southern all-sky survey \\
       & 850  &  7.3 & 85 & 3.69  & 1.0     & Southern all-sky survey \\\hline
KIDS-3 & 1000 &  6.2 & 50 & 40.7  & 0.65             & -- \\
       & 1300 &  4.8 & 65 & 38.4  & 0.04             & -- \\
       & 1500 &  4.1 & 75 & 81.6  & 0.04             & -- \\\hline
KIDS-4 & (2000)\footnotemark[$\ddag$] & (3.1)\footnotemark[$\ddag$] & (50)\footnotemark[$\ddag$] & (716.2)\footnotemark[$\ddag$] & (0.005)\footnotemark[$\ddag$] & Planned values \\
\hline
\end{tabular}
}
\begin{tabnote}
  \begin{minipage}[t]{0.9\textwidth}
  \raggedright
  \footnotemark[$*$] KIDS-1/2 are optimized for the southern all-sky survey; KIDS-3/4 for higher-frequency follow-up. Planned FoV $>$1~deg$^2$.\\
\footnotemark[$\dag$]
Sensitivities were computed by including all noise contributions from the 
atmosphere, telescope, and detectors (i.e., sky + antenna + detector background), 
using equation (\ref{eq:app1_A4}) and parameters in Appendix~\ref{sec:app1}. 
Longer integrations scale as $t^{-1/2}$.
\\
  \footnotemark[$\ddag$] Parentheses denote tentative design goals.
  \end{minipage}
\end{tabnote}
\label{tab:kids_specs}
\end{table*}

\begin{table*}[t]
\tbl{Comparison of representative wide-field (sub-)mm surveys.}{%
\centering
\begin{tabular}{lcccc}
\hline
Survey & Telescope/Instrument & Wavelength & Area [deg$^2$] & Depth (5$\sigma$)\footnotemark[$*$] \\
\hline\hline
ATT12 (KIDS-1/2) & 12-m + KIDS (MKIDs) & 1000, 750, 600, 460, 350 $\mu$m & $\sim$10,000 & Confusion-limited ($\sim$1.0--1.7 mJy) \\
ATT12 (KIDS-3/4) & 12-m + KIDS (MKIDs) & $\sim$200--300 $\mu$m (higher-$\nu$ follow-up) & Targeted & Shallower sensitivity than KIDS-1/2\\
SPT-SZ/SPTpol\footnotemark[$a$] & 10-m SPT & 1400, 2000, 3200 $\mu$m & $\sim$2500 & $\sim$6 mJy (at 2000 $\mu$m) \\
H-ATLAS\footnotemark[$b$] & \textit{Herschel}/SPIRE & 250, 350, 500 $\mu$m & $\sim$600 & 32--145 mJy \\
HerMES\footnotemark[$c$] & \textit{Herschel}/PACS+SPIRE & 100--500 $\mu$m & $\sim$380 & $\sim$29, 32, 34 mJy (SPIRE) \\
HeLMS\footnotemark[$d$] & \textit{Herschel}/SPIRE & 250--500 $\mu$m & $\sim$270 & $\sim$52 mJy (at 500 $\mu$m) \\
HerS\footnotemark[$e$] & \textit{Herschel}/SPIRE & 250--500 $\mu$m & $\sim$79 & $\sim$65, 64, 74 mJy \\
PEP\footnotemark[$f$] & \textit{Herschel}/PACS & 100, 160 $\mu$m & $\sim$10--20 & $\sim$6--8 mJy (100), $\sim$12--17 mJy (160) \\
GOODS-Herschel\footnotemark[$f$] & \textit{Herschel}/PACS+SPIRE & 100--500 $\mu$m & $<$1 & $\sim$1--2 mJy (PACS deep) \\
\hline
\end{tabular}
}
\begin{tabnote}
  \begin{minipage}[t]{0.9\textwidth}
  \raggedright
  \footnotemark[$*$] Approximate 5$\sigma$ sensitivities from the cited literature.\\
  \footnotemark[$a$]\citet{Carlstrom2011,Reuter2020};\,
  \footnotemark[$b$]\citet{Eales2010};\,
  \footnotemark[$c$]\citet{Oliver2012};\,
  \footnotemark[$d$]\citet{Asboth2016};\,
  \footnotemark[$e$]\citet{Viero2014};\,
  \footnotemark[$f$]\citet{Lutz2011,Elbaz2011}.
  \end{minipage}
\end{tabnote}
\label{tab:survey_comparison}
\end{table*}

\begin{table*}[t]
\tbl{Representative rest-frame FIR and THz emission lines accessible to ATT12 within its spectroscopic frequency coverage.}{%
\centering
\begin{tabular}{lccc}
\hline
Line & Frequency (THz)\footnotemark[$*$] & Transition & $E_{\rm u}/k$ [K]\footnotemark[$\dag$] \\
\hline\hline
CO ($J=$ 9--8 to 13--12) & 1.037--1.497 & rotational & 249--503 \\
CH 205~\micron & 1.471 & -- & 96 \\
CH$^{+}$ (1--0) 359~\micron & 0.835 & $J=1 \rightarrow 0$ & 40 \\
CH$^{+}$ (2--1) 180~\micron & 1.669 & $J=2 \rightarrow 1$ & 120 \\
OH$^{+}$ $1(1)$--$0(1)$ 290~\micron & 1.033 & rotational & 50 \\
H$_{2}$O (multiple) 150--1500~\micron & 0.2--2.0 & various rotational & 100--500 \\
H$_{2}$D$^{+}$ 219~\micron & 1.370 & ortho $1_{1,0}$--$1_{1,1}$ & 66 \\
{[C\,{\sc i}]} (1--0) 609~\micron & 0.492 & $^3P_{1} \rightarrow ^3P_{0}$ & 24 \\
{[C\,{\sc i}]} (2--1) 370~\micron & 0.809 & $^3P_{2} \rightarrow ^3P_{1}$ & 62 \\
{[C\,{\sc ii}]} 158~\micron & 1.900 & $^2P_{3/2} \rightarrow ^2P_{1/2}$ & 91 \\
{[O\,{\sc i}]} 63~\micron & 4.744 & $^3P_{1} \rightarrow ^3P_{2}$ & 228 \\
{[O\,{\sc i}]} 145~\micron & 2.060 & $^3P_{0} \rightarrow ^3P_{1}$ & 327 \\
{[O\,{\sc iii}]} 88~\micron & 3.393 & $^3P_{1} \rightarrow ^3P_{0}$ & 163 \\
{[O\,{\sc iii}]} 52~\micron & 5.786 & $^3P_{2} \rightarrow ^3P_{1}$ & 326 \\
{[N\,{\sc ii}]} 205~\micron & 1.461 & $^3P_{1} \rightarrow ^3P_{0}$ & 70 \\
{[N\,{\sc ii}]} 122~\micron & 2.461 & $^3P_{2} \rightarrow ^3P_{1}$ & 188 \\
{[N\,{\sc iii}]} 57~\micron & 5.259 & $^2P_{3/2} \rightarrow ^2P_{1/2}$ & 251 \\
\hline
\end{tabular}
}
\begin{tabnote}
  \begin{minipage}[t]{0.9\textwidth}
  \raggedright
  \footnotemark[$*$] Rest-frame frequencies; corresponding wavelengths shown for clarity.\\ Several short-wavelength lines exceed 2~THz and enter the ATT12 bands only when redshifted to 275--2000~GHz.\\
  \footnotemark[$\dag$] Upper-level energies $E_{\rm u}/k$ when available.
  \end{minipage}
\end{tabnote}
\label{tab:Thz-line}
\end{table*}

In spectroscopy, ATT12 covers 275--2000~GHz, providing access to key molecular THz transitions (e.g., high-$J$ CO, CH, H$_2$D$^+$) in star-forming regions \citep[e.g.,][]{Hirashita2016}, and FIR fine-structure lines in nearby galaxies \citep[e.g.,][]{DeLooze2014,Peng2025a,Peng2025b,Peng2025c}.
Representative transitions are summarized in table~\ref{tab:Thz-line}.
At high redshift, these lines redshift into high-transmission windows, enabling efficient observations.

\subsection{Organization of the Paper}
\label{subsec:intro3}

We investigate ATT12’s capabilities, focusing first on heterodyne spectroscopy and then on wide-field imaging with the MKID camera.
We assess detectability of FIR fine-structure lines from DSFGs using atmospheric/sensitivity simulations under typical Dome Fuji winter PWV, deriving detection limits, implied $L_{\rm IR}$, and accessible redshift ranges.
We then estimate the statistical yield of confusion-limited continuum surveys (source counts and basic properties) and discuss how diagnostic line ratios constrain ISM conditions, linking spectroscopy and continuum views.

The paper is organized as follows:
section~\ref{sec:2} describes assumptions, instrumental parameters, and simulations;
section~\ref{sec:3} presents detectability of key FIR lines and implications for galaxy properties;
in section~\ref{sec:4} we evaluate source counts and characteristics from wide-field continuum surveys;
section~\ref{sec:5} summarizes our conclusions.
We adopt a flat $\Lambda$CDM cosmology consistent with \citet{Planck2020} with $H_0=67.4$~km~s$^{-1}$~Mpc$^{-1}$, $\Omega_{\mathrm{m}}=0.315$, and $\Omega_{\Lambda}=0.685$, and a solar luminosity of $L_\odot=3.842\times10^{33}$~erg~s$^{-1}$.

\section{Spectroscopic Sensitivity of the Antarctic 12-m Terahertz Telescope (ATT12)}
\label{sec:2}

Although wide-field imaging surveys are expected to provide the initial galaxy samples (see section~\ref{subsec:intro3}), 
we present the spectroscopic sensitivity first, followed by the imaging survey performance in section~\ref{subsec:3.3}. 
This choice reflects the primary focus of the present study on the feasibility of FIR line spectroscopy with ATT12. 
Here we describe the methodology used to calculate the spectroscopic sensitivity of ATT12 as a function of frequency.

\subsection{Atmospheric optical depth at ATT12}
\label{subsec:2.1}

We evaluate the atmospheric optical depth, $\tau$, at the planned site of ATT12 (see also \citealt{Matsuo2019}). 
The calculations were performed using the atmospheric transmission model ({\tt am}; \citealt{Paine2019}), which solves the radiative transfer equations from the microwave to the submillimeter regime. 
The model treats the atmosphere as a stack of plane-parallel layers in hydrostatic equilibrium, each characterized by temperature, pressure, and molecular abundances specified by the user.

In {\tt am}, $\tau$ is computed by integrating molecular absorption over altitude. 
For water vapor (H$_2$O), both line and continuum absorption are included. 
The latter arises from self-collisions (H$_2$O--H$_2$O) and collisions with other molecules such as N$_2$ and O$_2$, and is modeled using the empirical MT\_CKD scheme \citep{Mlawer2023}. 
In this work, we adopt the January~2023 release of the MT\_CKD model\footnote{\url{https://github.com/AER-RC/MT_CKD}} for continuum absorption, and include the main line absorbers H$_2$O and O$_3$.

Figure~\ref{fig:atm_tau} shows the resulting $\tau$ spectra between 100 and 2000~GHz with a frequency resolution of 10~MHz, assuming a zenith angle of $40^{\circ}$. 
The input profiles of temperature, pressure, and molecular column densities are based on those measured at Dome Fuji, the closest well-characterized site. 
Since New Dome Fuji is expected to have lower PWV, we adopt values representative of Dome~A, which exhibits drier and more stable conditions. 
Specifically, the 25th- and 50th-percentile winter PWV values at Dome~A (100 and 150~$\mu$m; \citealt{H.Yang2010}) are used as proxies for New Dome Fuji, corresponding to the ``winter~25\%'' and ``winter~50\%'' conditions, respectively (see also section~\ref{subsec:intro2}). 

We define ``atmospheric windows'' as frequency intervals where $\tau \lesssim 3.0$, corresponding to transmission $\gtrsim 5$\%, following \citet{Oberst2006}. 
As shown in figure~\ref{fig:atm_tau}, New Dome Fuji is expected to provide excellent transparency with $\tau \leq 0.6$ up to 850~GHz under both PWV conditions. 
In the ``winter~25\%'' case (PWV = 100~$\mu$m), additional windows appear at 
1000.0--1059.6~GHz, 1246.0--1390.5~GHz, 1432.1--1579.2~GHz, 1821.9--1838.5~GHz, and 1952.7--2000.0~GHz.

\begin{figure}[t]
\centering
\includegraphics[width=0.45\textwidth]{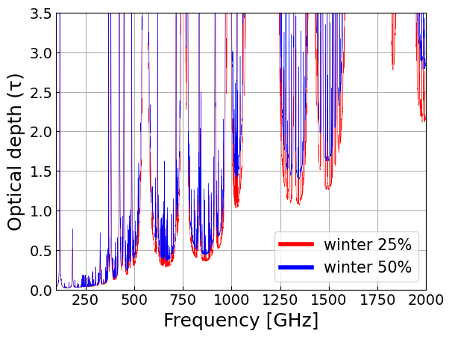}
\caption{\raggedright
Atmospheric optical depth ($\tau$) at a zenith angle of 40$^{\circ}$ as a function of frequency at New Dome Fuji, 
simulated using the {\tt am} model. 
The red and blue curves correspond to the 25th- and 50th-percentile winter PWV conditions at Dome~A 
(100 and 150~$\mu$m, respectively), assumed to approximate those at New Dome Fuji. 
}
\label{fig:atm_tau}
\end{figure}

\subsection{System Noise Temperature of ATT12}
\label{subsec:2.2}

ATT12 will be equipped with a multi-beam heterodyne receiver system for spectroscopic observations. 
Here we describe the assumptions adopted for estimating the system noise temperature and resulting sensitivity. 
Spectroscopic follow-up observations are assumed for sources with photometric redshifts derived from the imaging surveys; 
details of the color–color selection will be presented in a forthcoming paper (Paper~II).

Our primary targets are distant DSFGs, including Ultraluminous infrared galaxies (ULIRGs), Hyperluminous infrared galaxies (HyLIRGs), and SMGs.
These galaxies are effectively point-like for ATT12 (see section~\ref{subsec:intro2}), with typical FIR half-light radii of 
$r_{\rm e}\sim0\farcs1$–$0\farcs5$ ($\sim$1–2 kpc) for $\log(L_{\rm IR}/L_\odot)\sim12$–13 \citep[e.g.,][]{Fujimoto2017}. 
Very nearby extended galaxies ($z\le0.01$) are not considered in this study.

We adopt the ON--ON observing method, in which two beams alternate between ON-source and OFF-source positions. 
This beam-switching mode, requiring at least two beams, reduces the effective observing time compared to a conventional single-beam ON--OFF method.

\begin{table}[t]
\tbl{Receiver and system noise temperatures at representative ATT12 observing frequencies.}{%
\centering
\begin{tabular}{cccc}
\hline 
Frequency & $T_{\text{RX}}$\footnotemark[$*$] & $T_{\text{sys}}$ (25\%)\footnotemark[$\dag$] & $T_{\text{sys}}$ (50\%)\footnotemark[$\dag$] \\ 
(GHz) & (K) & (K) & (K) \\ 
\hline\hline
600  & 120 & 490   & 700   \\ 
850  & 160 & 370   & 420   \\ 
1050 & 350 & 2410  & 4340  \\ 
1500 & 500 & 2430  & 3560  \\
2000 & 670 & 8010  & 16290 \\ 
\hline
\end{tabular}
}
\begin{tabnote}
  \begin{minipage}[t]{0.45\textwidth}
  \raggedright
  \footnotemark[$*$] Receiver temperature.\\
  \footnotemark[$\dag$] System temperatures for ``winter 25\%'' and ``winter 50\%'' PWV conditions at a zenith angle of $40^\circ$.
  \end{minipage}
\end{tabnote}
\label{table:Trx_Tsys}
\end{table}

The noise-equivalent sensitivity $\Delta T$ of a beam-switching single-dish system with a single-polarization receiver is
\begin{equation}
\Delta T = \frac{\sqrt{2}\,T_{\mathrm{sys}}}{\sqrt{B\,t}},
\end{equation}
where $T_{\mathrm{sys}}$ is the system noise temperature, $B$ the spectral bandwidth, and $t$ the on-source integration time.

The system noise temperature is calculated as
\begin{align}
T_{\mathrm{sys}} = \frac{1}{\eta}
\bigg\{ T_{\mathrm{RX}} 
&+ \eta\, T_{\mathrm{atm}} \left( 1 - e^{-\tau \sec Z} \right)
+ (1-\eta)\,T_{\mathrm{amb}} \bigg\}
e^{\tau \sec Z},
\end{align}
where $T_{\mathrm{RX}}$ is the receiver temperature, $\tau$ the atmospheric optical depth, $Z$ the zenith angle, and $\eta$ the feed efficiency (fixed to 0.95).
The atmospheric temperature is assumed to equal the winter ambient temperature, $T_{\mathrm{amb}}=208$~K.

The receiver operates in single-sideband (SSB) mode below $1$~THz and in double-sideband (DSB) mode above $1$~THz. 
We assume SIS receivers with noise temperatures of four times the quantum limit below 1~THz, and seven times the limit above 1~THz 
\citep{fujii2016low,uzawa2009sensitive,rudakov2021thz}. 
The Rayleigh–Jeans equivalent quantum noise temperature is
\begin{equation}
T = \frac{h\nu}{k_B},
\end{equation}
where $\nu$ is the observing frequency.

Table~\ref{table:Trx_Tsys} summarizes the adopted quantum noise, receiver noise, and system temperatures. 
These values represent conservative assumptions; future technical advances may further reduce them. 
At high frequencies, atmospheric transmission becomes the dominant factor influencing $T_{\mathrm{sys}}$.

\subsection{Observability of FIR Fine-Structure Lines with ATT12}
\label{subsec:2.3}

\begin{figure*}[ht]
    \centering
    \includegraphics[width=0.48\linewidth]{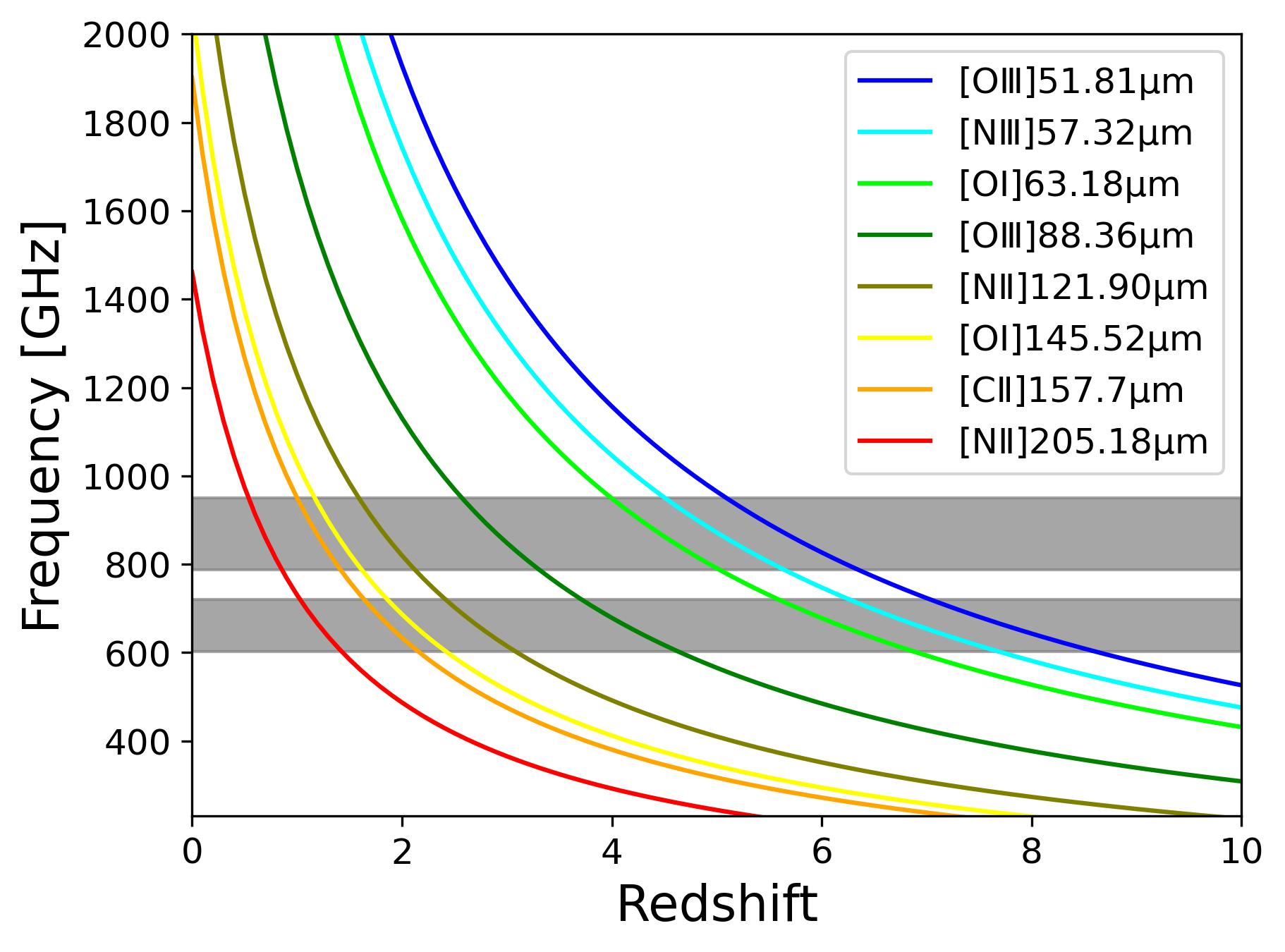}
    \includegraphics[width=0.48\linewidth]{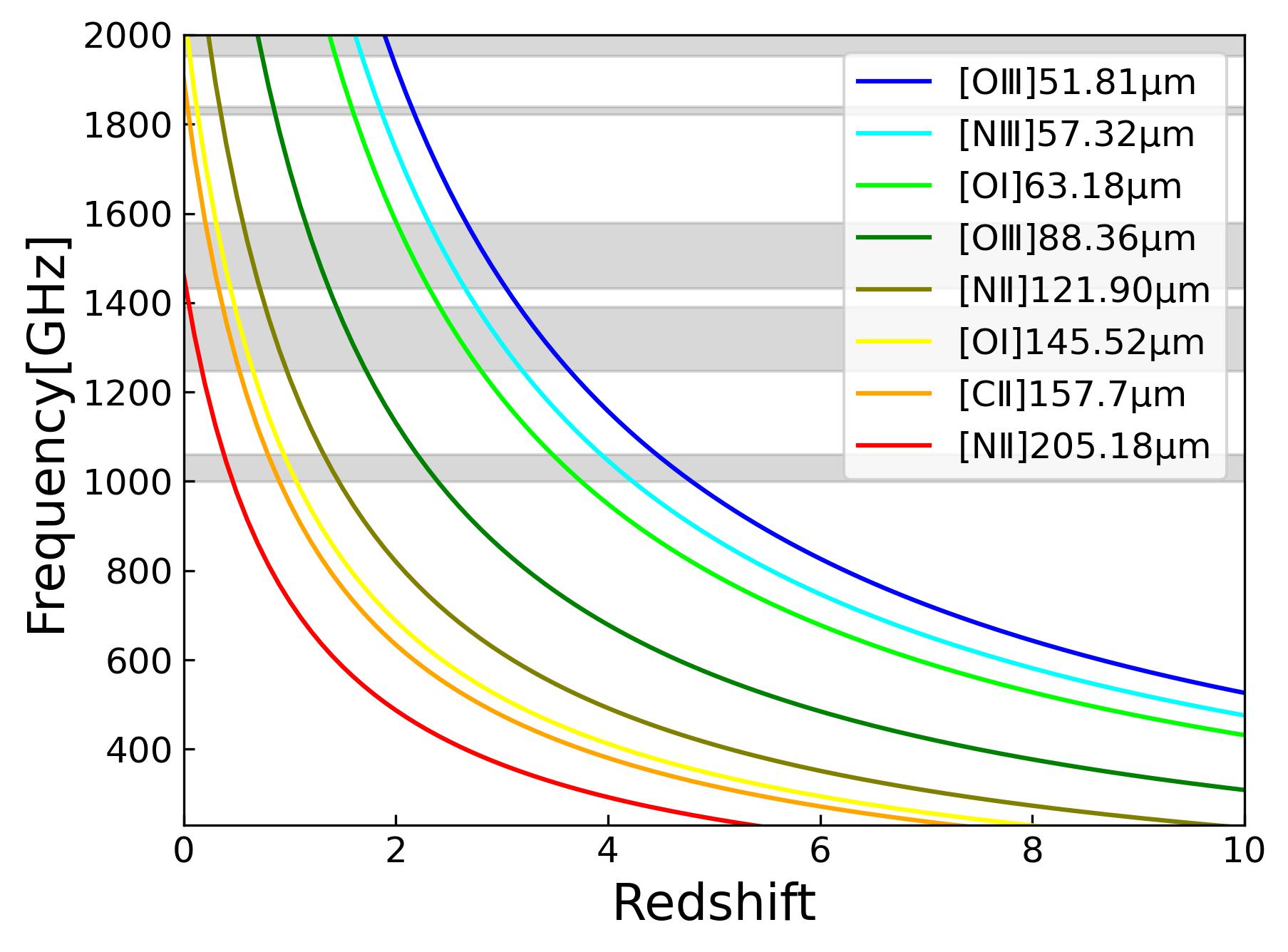}
    \caption{\raggedright
    Redshift–frequency relation for major FIR fine-structure lines. 
    The left panel marks ALMA Bands~9 (lower) and 10 (upper), while the right panel highlights
    the atmospheric windows under the ``winter 25\%'' PWV condition (100~$\mu$m). 
    }
    \label{fig:line_redshift}
\end{figure*}

Using the atmospheric windows derived in section~\ref{subsec:2.1}, we assess the detectability of key FIR fine-structure lines introduced in section~\ref{subsec:intro1}. 
The left panel of Figure~\ref{fig:line_redshift} illustrates the limited accessibility of ALMA Bands~9 and 10, which require exceptional atmospheric conditions at the Atacama site (e.g., PWV $\lesssim0.3$ mm for Band~10). 
Such conditions occur only during the best $\sim$5\% of the time at Chajnantor \citep{Radford2016}. 
In contrast, at New Dome Fuji PWV $\lesssim0.3$ mm is expected to occur for $\gtrsim50\%$ of the time \citep{H.Yang2010,Matsuo2019}. 
The substantially higher atmospheric transparency therefore allows ATT12 to operate efficiently across these frequency ranges and to access the THz windows for a much larger fraction of the time.

Among the eight brightest FIR lines \citep[e.g.,][]{Peng2025b, Decarli2025},  
ATT12 accesses long-wavelength lines 
(\nii~122~$\mu$m, \oi~145~$\mu$m, \cii~158~$\mu$m, \nii~205~$\mu$m) 
from $z\sim0$–2,  
and short-wavelength lines 
(\oiii~52~$\mu$m, \niii~57~$\mu$m, \oi~63~$\mu$m, \oiii~88~$\mu$m) 
primarily at $z\sim1$–5.  
Atmospheric absorption, however, produces gaps in redshift coverage.

We next compare ATT12 with other facilities.  
Its heterodyne frequency range (275–2000~GHz) bridges the gap between ALMA (up to 950~GHz) and the space-based FIR coverage of PRIMA/FIRESS (24–235~$\mu$m; 1.28–12.5~THz).  
PRIMA will deliver continuous FIR spectroscopy with $\sim$an order-of-magnitude sensitivity improvement over \textit{Herschel}, albeit with modest spatial resolution (e.g., $\sim$22\arcsec at 158~$\mu$m), which is significantly coarser than ATT12’s few-arcsecond beams.  
Thus, ATT12 will mitigate source confusion thanks to its higher spatial resolution, while PRIMA, although having coarser beams, may nevertheless avoid confusion through its hyperspectral imaging capability \citep[e.g.,][]{Bethermin2024}, and will provide access to FIR frequencies inaccessible from the ground.

Compared to ALMA, ATT12 will have lower line sensitivity, reflecting the difference in collecting area.  
However, ATT12 recovers spatially extended emission that ALMA may resolve out, and its multi-beam mapping allows efficient large-area surveys.  
Relative to SOFIA (2.5~m), whose heterodyne instruments covered several 0.5–4.7~THz windows before decommissioning, ATT12 offers substantially smaller beams at comparable frequencies.

These complementarities motivate coordinated use: ATT12 provides high-resolution THz mapping and access to atmospheric windows unavailable to ALMA, while PRIMA supplies ultra-sensitive FIR spectroscopy from space.  
Table~\ref{tab:facility_comparison} summarizes the capabilities of the relevant observatories.

\begin{table*}[t]
\tbl{Frequency/wavelength coverage, apertures, and representative beams for facilities discussed in this work.}{%
\centering
\begin{tabular}{lccc}
\hline
Facility & Spectral coverage & Primary mirror $D$ & Representative beam\footnotemark[$*$] \\
\hline\hline
ATT12 (this work) & 275--2000~GHz ($\lambda\!\approx$1.5--0.15 mm) & 12~m & $\sim$6\farcs3 at 1 THz; $\sim$3\farcs1 at 2 THz \\
ALMA & 35--950~GHz (Bands 1--10) & 12~m (array) & 19\arcsec primary beam at 300~GHz (single 12~m) \\
Herschel/PACS & 70, 100, 160~$\mu$m & 3.5~m & $\sim$5\farcs6--11\farcs3 \\
Herschel/SPIRE & 250, 350, 500~$\mu$m; FTS 200--670~$\mu$m & 3.5~m & $\sim$18\farcs2--36\farcs3 \\
PRIMA/FIRESS & 24--235~$\mu$m (1.28--12.5 THz) & 1.8~m & $\sim$22\arcsec at 158~$\mu$m; $\sim$12\farcs3 at 88~$\mu$m \\
SPT (SPT-3G) & 95, 150, 220~GHz & 10~m & $\sim$1\arcmin at 150 GHz \\
\hline
\end{tabular}
}
\begin{tabnote}
  \begin{minipage}[t]{0.9\textwidth}
  \raggedright
  \footnotemark[$*$] Diffraction-limited beams estimated using $\theta_{\rm HPBW}=1.2\,\lambda/D$.  
  ALMA value corresponds to a single 12-m antenna; synthesized resolution depends on configuration.
  \end{minipage}
\end{tabnote}
\label{tab:facility_comparison}
\end{table*}

\section{Results}
\label{sec:3}

We now present the results of our feasibility analysis. 
Sections~\ref{subsec:3.1} and \ref{subsec:3.2} focus on the spectroscopic sensitivity of ATT12, quantifying the detection limits for individual fine-structure lines and the corresponding infrared luminosities. 
Section~\ref{subsec:3.3} then turns to the expected number of galaxies detectable in the wide-field imaging surveys with the MKIDs camera.

\subsection{Detection Limits for Fine-Structure Lines}
\label{subsec:3.1}

\begin{figure*}[t]
    \centering
    \includegraphics[width=0.9\linewidth]{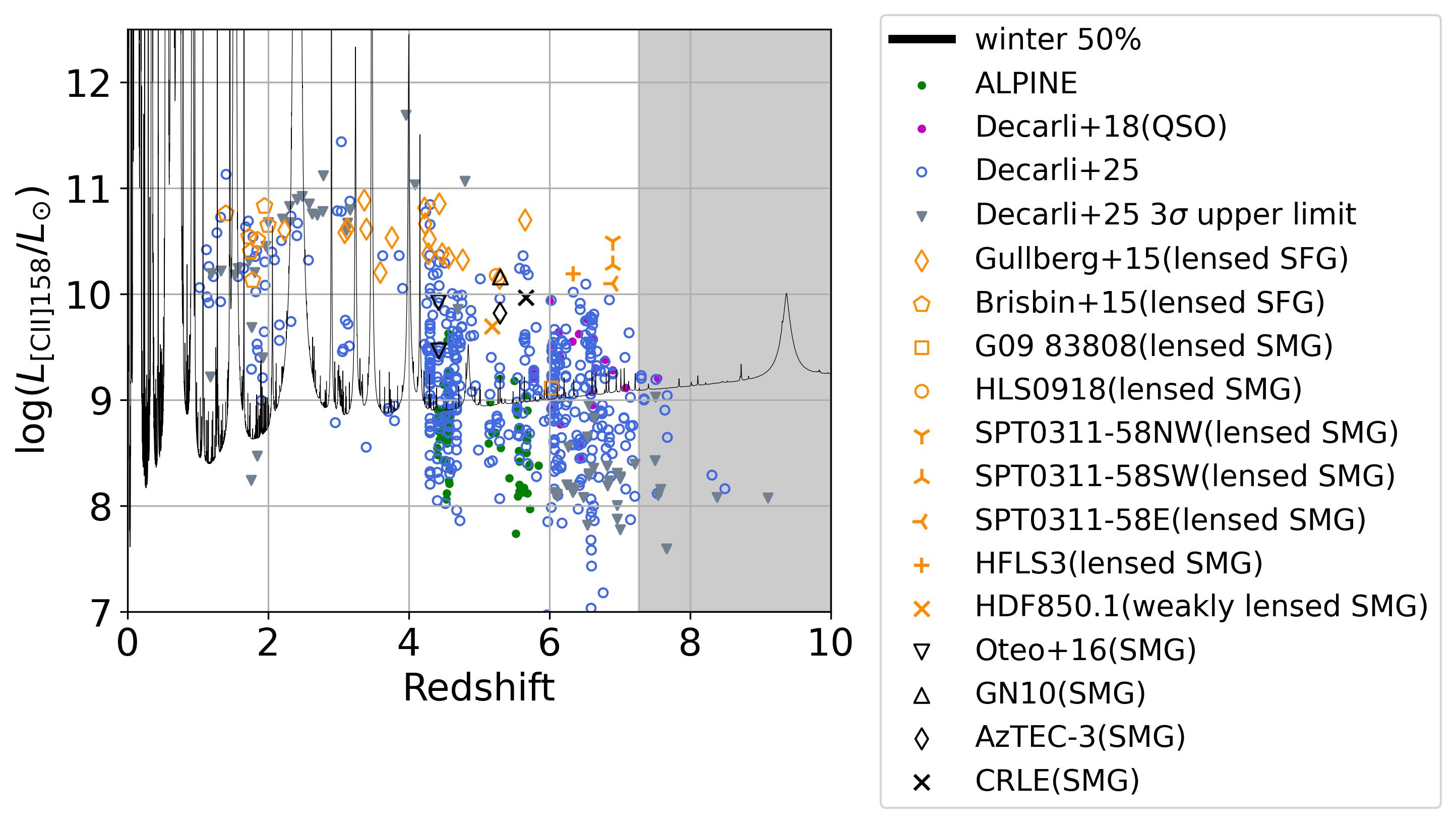}
    \caption{\raggedright
    [C\,\textsc{ii}] 158\,$\mu$m line luminosity vs.\ redshift. 
    The black curve indicates the ATT12 5$\sigma$ detection limit for an effective observation time of 100 hours per source under the ``winter 50\%'' condition. The shaded regions indicate frequencies outside ATT12's operational range.
    Observed [C\,\textsc{ii}] luminosities from ALPINE \citep{Bethermin2020}, quasars \citep{Decarli2018}, and SMGs \citep{Combes2012, Walter2012, Gullberg2015, Brisbin2015, Oteo2016, Pavesi2016, Zavala2018, riechers2013dust, Litke2023, marrone2018galaxy} are overplotted. Green circles indicate Lyman-break galaxies from the ALPINE program \citep{Bethermin2020}, purple circles denote quasars \citep{Decarli2018}, and orange and black symbols represent lensed and unlensed SMGs, respectively. We also include the compilation of \cite{Decarli2025}, excluding sources already listed in the aforementioned studies. The luminosities are not corrected for gravitational lensing. 
    } \par
    \label{fig:CII_limit}
    
\end{figure*}

\begin{figure*}[!ht]
    \centering
    \includegraphics[width=0.49\linewidth]{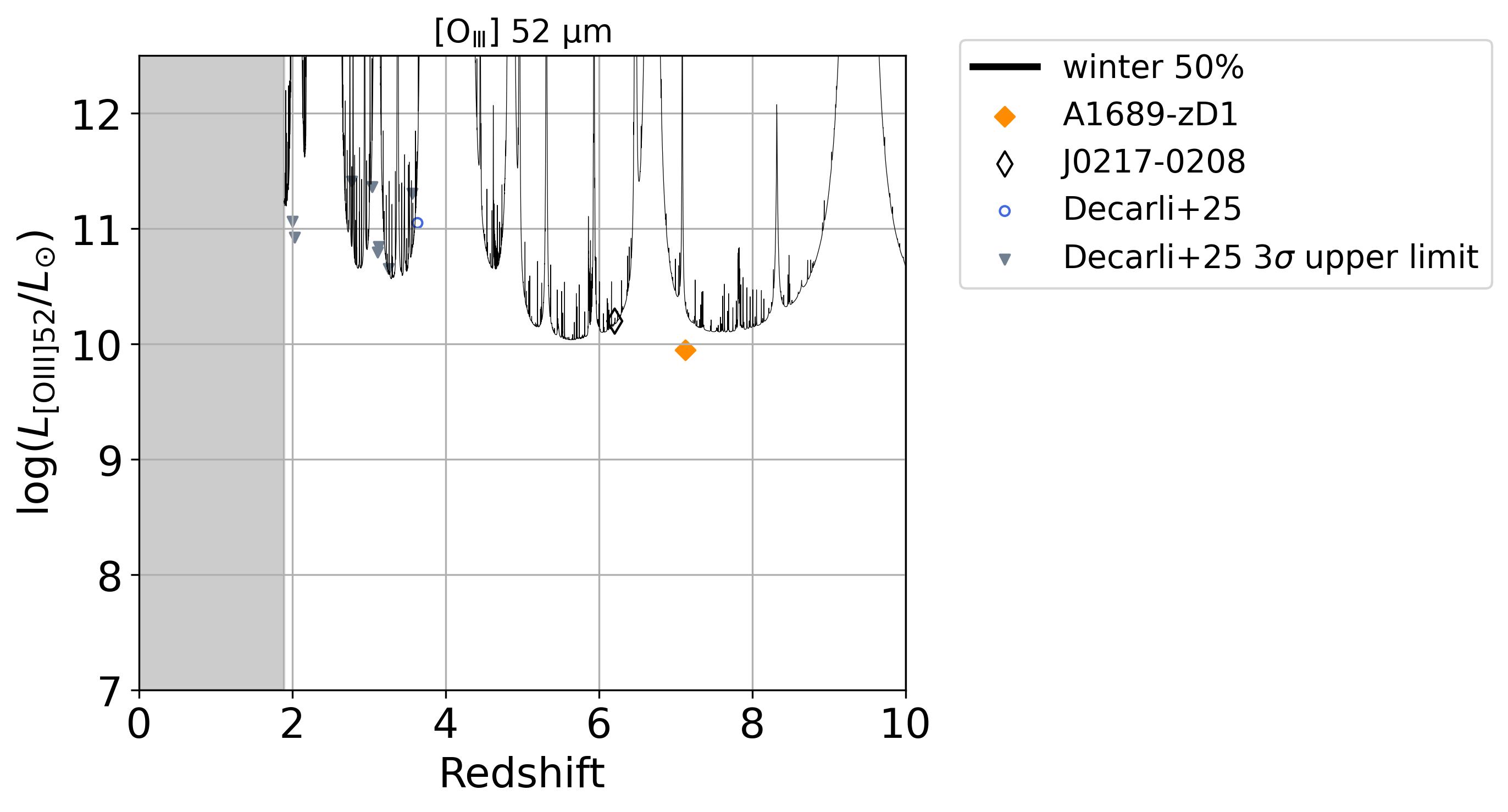}
    \includegraphics[width=0.42\linewidth]{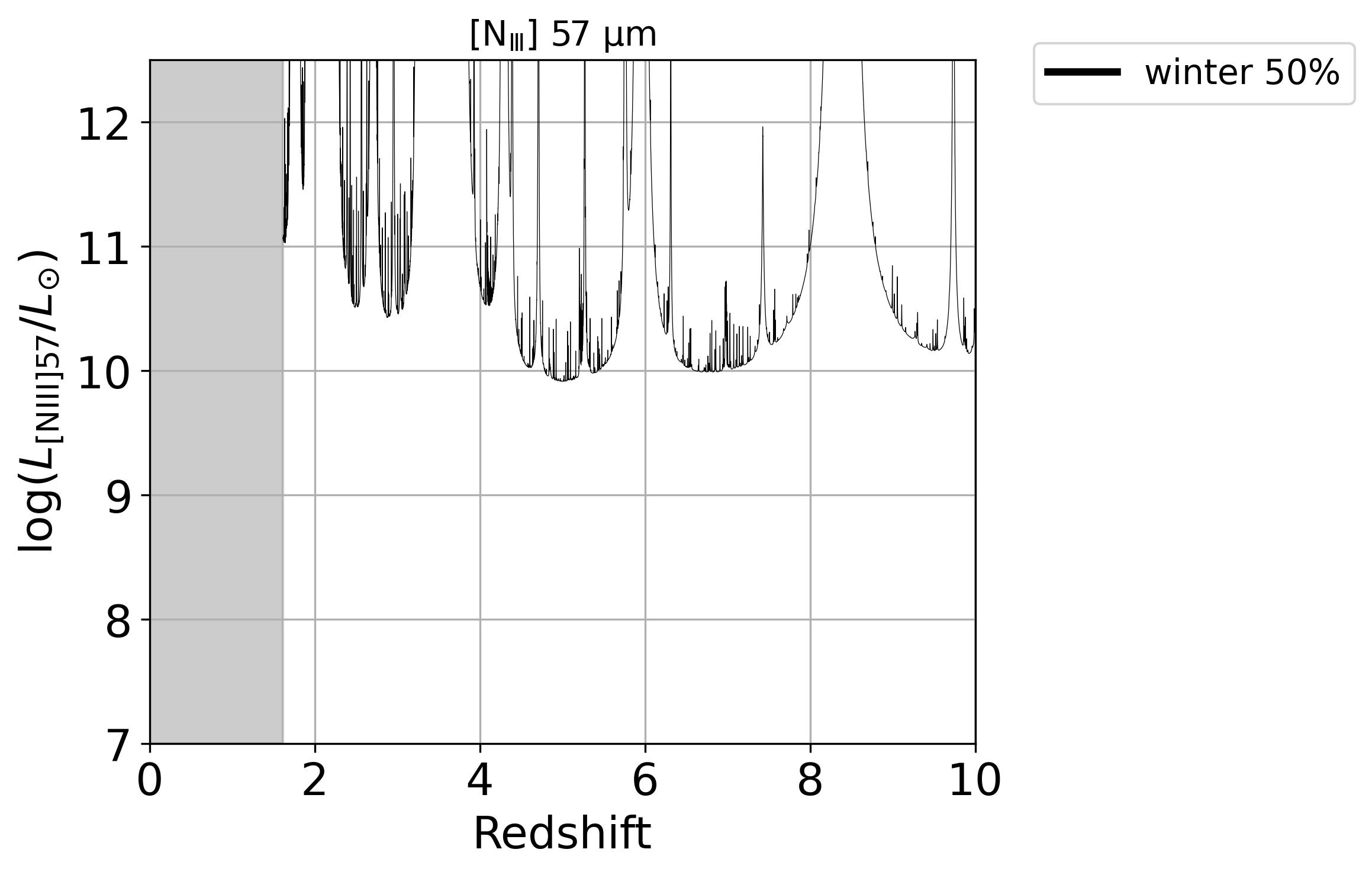}
    \includegraphics[width=0.49\linewidth]{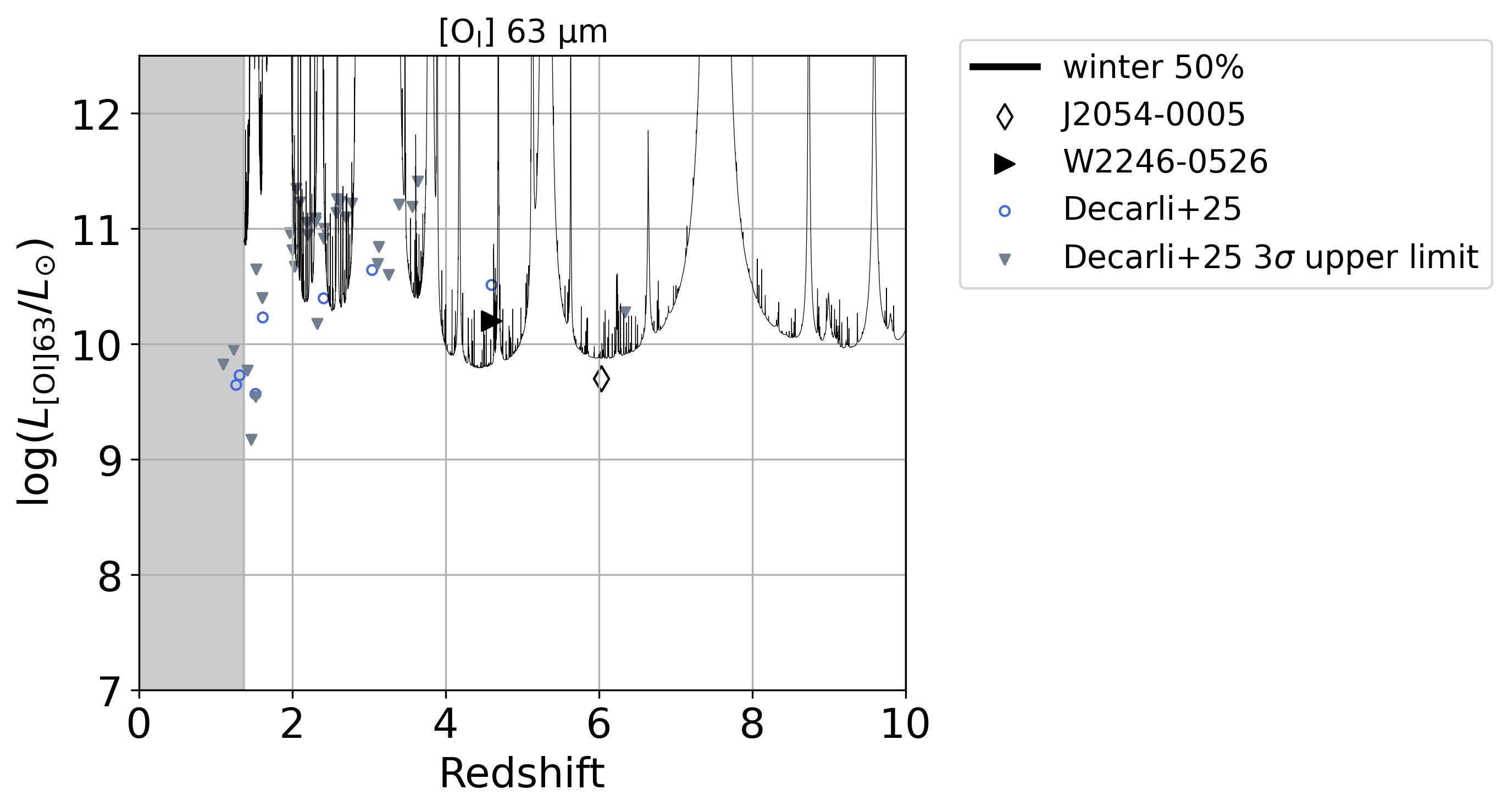}
    \includegraphics[width=0.49\linewidth]{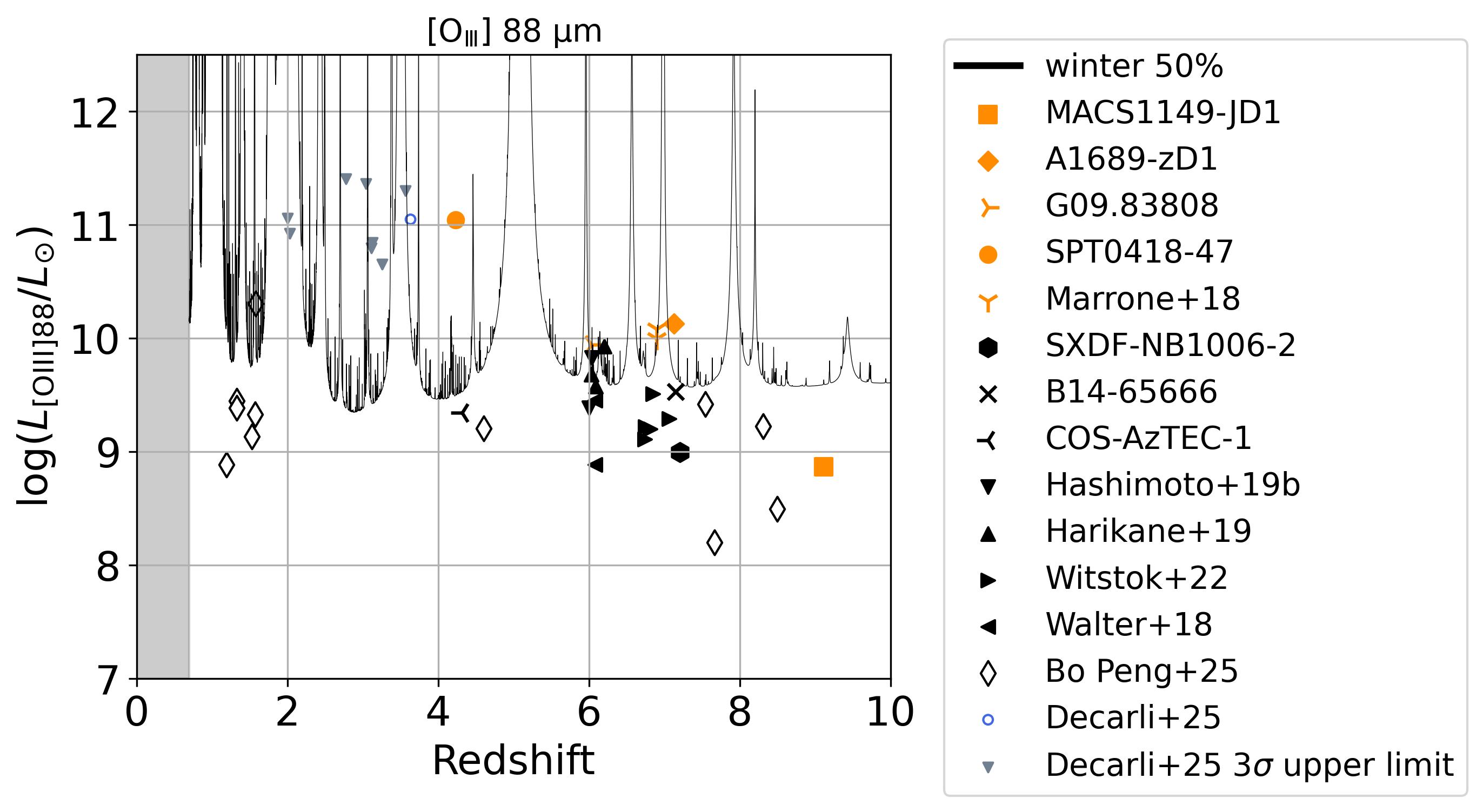}
    \includegraphics[width=0.49\linewidth]{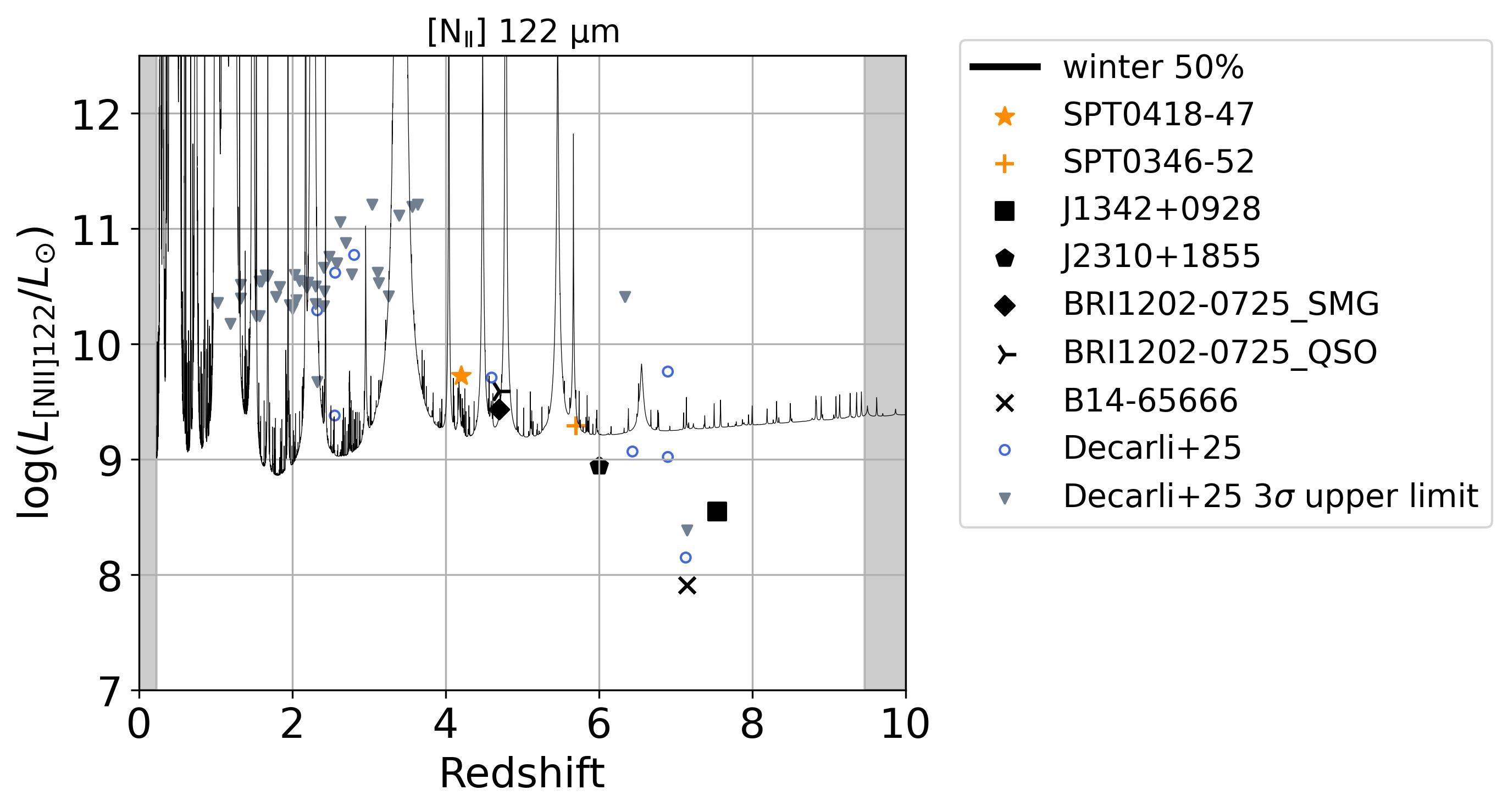}
    \includegraphics[width=0.49\linewidth]{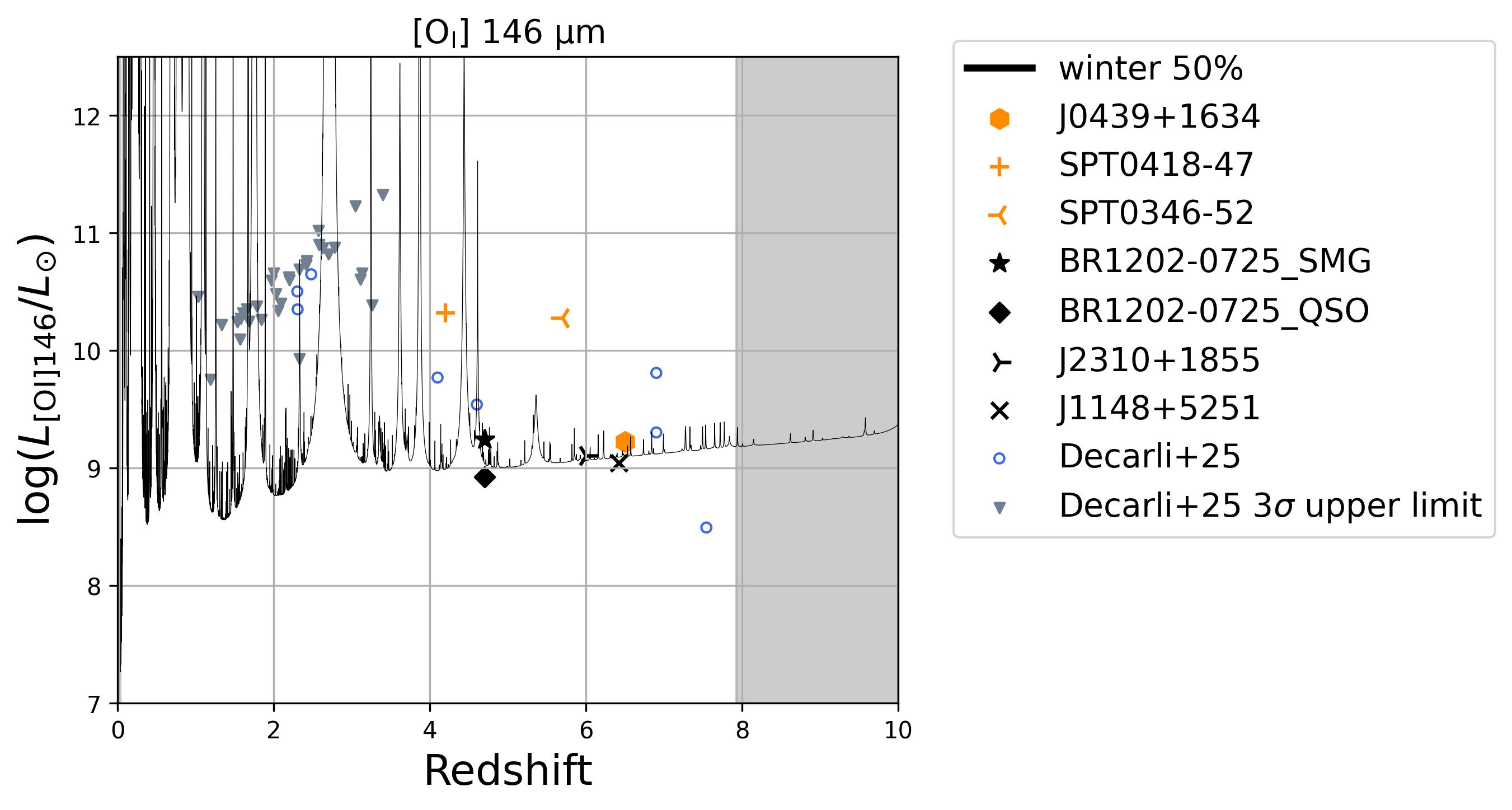}
    \includegraphics[width=0.49\linewidth]{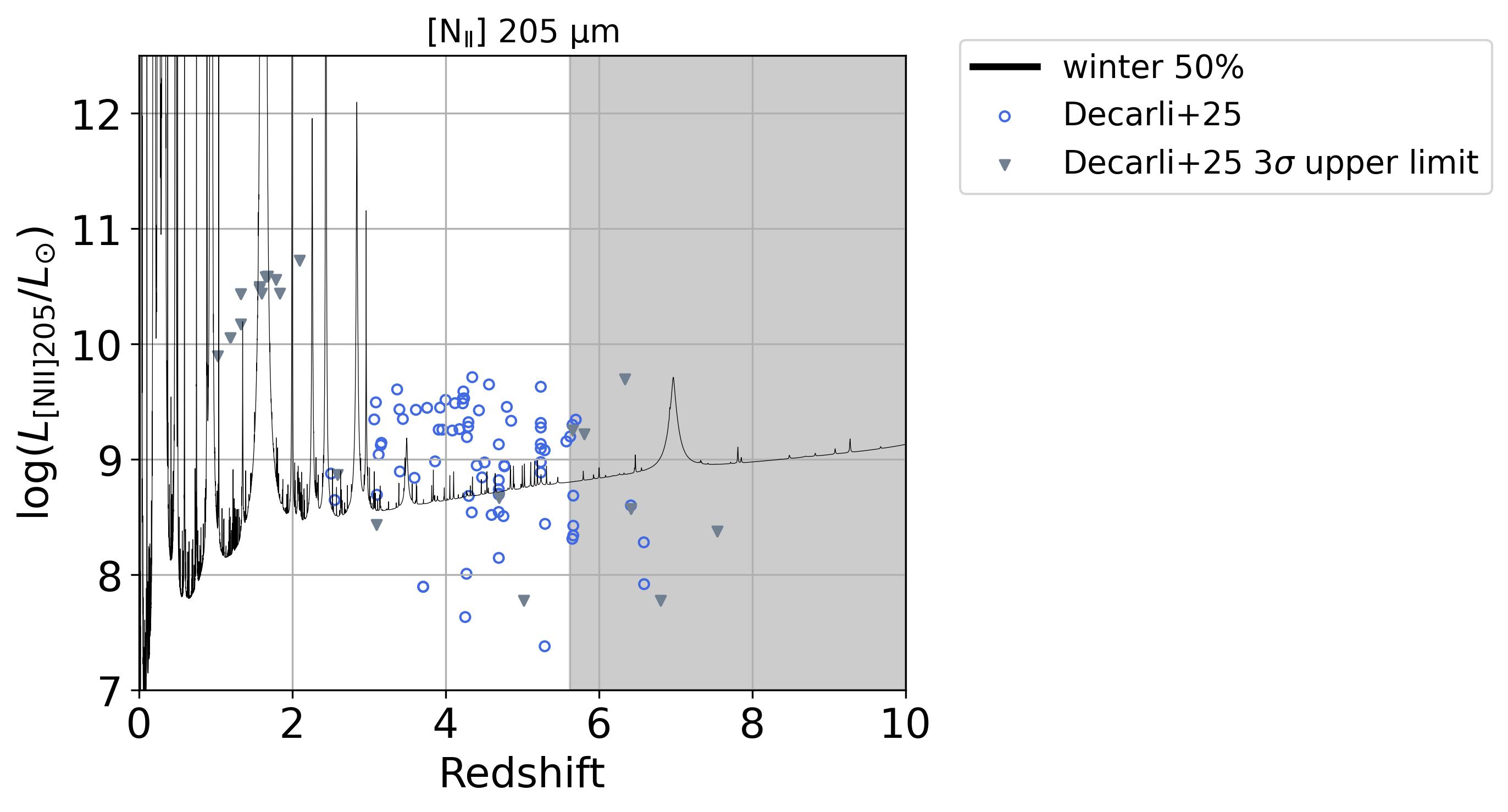}
    \caption{\raggedright
    Same as figure~\ref{fig:CII_limit}, but for the other seven brightest FIR emission lines. 
    }\par
    \label{fig:Great8_limit}
\end{figure*}

We estimate the minimum detectable flux density for FIR fine-structure lines using the spectroscopic capability of ATT12. An ON–ON observation strategy is assumed, in which two beams alternate between the ON position. Each beam spends 50 hours on-source, yielding a total effective integration time of 100 hours per target. Given that ATT12 will allocate $\sim$3000 hours per year to scientific observations (consistent with the estimate in \S\ref{subsec:intro2}), such an investment per source is both realistic and feasible. The spectral bandwidth $B$ is set to correspond to a velocity width of 100 km s$^{-1}$, narrower than the typical line widths of $v \sim 300$ km s$^{-1}$ commonly observed in DSFGs (e.g., \citealt{Bothwell2013, marrone2018galaxy, Reuter2020, Bakx2020, Hagimoto2023}).

Using the noise temperature $\Delta T$ derived under these conditions (see eq.~(3) and table~\ref{table:Trx_Tsys}), the minimum detectable flux density $\Delta S$ for ATT12 is given by:
\begin{equation}
    \Delta S = \frac{2 k_\text{B}}{\eta_A A_\text{p}}\, \Delta T',
\end{equation}
where $\eta_{\rm A}A_\text{p}$ represents the effective antenna aperture area.

The detection criterion is defined such that the peak flux density of the observed line must be at least five times the noise level, i.e., $\Delta T' = 5\Delta T$, over one-third of the line width (100~km~s$^{-1}$).

The minimum detectable luminosity of a line can be expressed as follows \citep{Carilli2013}:
\begin{equation}
    L_\text{line} = 1.04 \times 10^{-3} \, \Delta S \, \Delta v \, D_\text{L}(z)^2 \, \nu_\text{obs} \quad [L_\odot],
\end{equation}
where $\Delta v$ is the spectral line velocity width in units of kilometers per second (km~s$^{-1}$), $D_\text{L}(z)$ is the luminosity distance at redshift $z$ in units of megaparsecs (Mpc), and $\nu_{\text{obs}}$ is the observed frequency in units of gigahertz (GHz). As noted above, we assume a line width of 300~km~s$^{-1}$.

Figure~\ref{fig:CII_limit} shows the 5$\sigma$ detection limit for the \cii~158~\micron\ line as a function of redshift. For comparison, observational data are overlaid from normal star-forming galaxies in the ALMA ALPINE survey at $z\approx4-6$ \citep{Le_Fevre2020, Bethermin2020}, $z > 5.94$ quasars \citep{Decarli2018}, and DSFGs \citep{Combes2012, Walter2012, Gullberg2015, Brisbin2015, Oteo2016, Pavesi2016} including the most distant sources such as G09 83808 \citep{Zavala2018}, HFLS3 \citep{riechers2013dust}, and SPT0311-58 \citep{Litke2023, marrone2018galaxy}. Lensed and unlensed sources are distinguished by orange and black symbols, respectively.  We also include the data compiled by \cite{Decarli2025}, after removing duplicated sources. 
This figure demonstrates that ATT12 will be capable of detecting DSFGs in \cii\ 158 \micron\ up to $z \sim 7$. In addition, ATT12 is expected to detect some bright normal star-forming galaxies and quasars.

Likewise, figure~\ref{fig:Great8_limit} shows the 5$\sigma$ detection limits for the other brightest FIR fine-structure lines with observational data where available: 
\oiii~52~$\mu$m, \niii~57~$\mu$m, \oi~63~$\mu$m, \oiii~88~$\mu$m, \nii~122~$\mu$m, \oi~145~$\mu$m, and \nii~205~$\mu$m. 

We first focus on \oiii~52~$\mu$m (the top-left panel) and \oi~63~$\mu$m (the left panel in the second row), for which recent ALMA Band~9 and~10 observations have reported detections in a small number of individual sources. These include a strongly lensed Lyman-break galaxy (LBG) A1689-zD1 at $z=7.13$ (\citealt{Killi2023}) and an intrinsically luminous LBG J0217$-$0208 at $z=6.20$ (\citealt{Harikane2025}) for \oiii~52~$\mu$m, as well as the hyper-luminous AGN W2246$-$0526 at $z=4.6$ (\citealt{fernandez_aranda2024}) and the FIR-luminous quasar J2054$-$0005 at $z=6.04$ (\citealt{Ishii2025}) for \oi~63~$\mu$m. These panels indicate that observations of these lines with ATT12 may require substantial integration times, although firmer conclusions will require larger and more statistically representative samples.

In the right panel of the second row of figure~\ref{fig:Great8_limit}, we include observational \oiii~88~$\mu$m data for quasars (\citealt{Walter2018, Hashimoto2019b, Novak2019}), DSFGs (\citealt{Tadaki2019, Tadaki2022, Debreuck2019, marrone2018galaxy}), and LBGs (\citealt{Inoue2016, Carniani2017, Hashimoto2018, Hashimoto2019, Tamura2019, Wong2022, Harikane2020, Witstok2022}), together with the recent compilation by \citet{Peng2025a} and \citet{Decarli2025}. 

The panels in the third row illustrate the observability of \nii~122~$\mu$m (left) and \oi~145~$\mu$m (right). In the left panel, we include observational data for three quasars, BR1202\_QSO, J1342+0928, and J2310+1855 (\citealt{Novak2019, Li2020, Lee2019}), as well as three DSFGs, SPT0418$-$47, SPT0346$-$52, and BR\_SMG (\citealt{De_breuck2019, Lee2019}), and one LBG, B14$-$65666 (\citealt{Sugahara2021}). We note that the data points for SPT0346$-$52 and B14$-$65666 represent $3\sigma$ upper limits. In the right panel, we include data for five quasars, BR1202\_QSO, J1342+0928, J2310+1855, J1148+5251, J04439+1634 (\citealt{Novak2019, Li2020, Lee2019, meyer2022, yang2019}) as well as three DSFGs (\citealt{De_breuck2019, Lee2019, litke2022}). 

Finally, in the panel in the fourth row, we include the compilation of \cite{Decarli2025} to illustrate the observability of \nii~205~$\mu$m. 
In these panels, we also include the compilation of \cite{Decarli2025}, excluding sources already listed in the aforementioned studies; $3\sigma$ upper limits for the non-detections are shown with downward triangles. These  panels show that ATT12 will be capable of detecting these lines in luminous DSFGs.

\subsection{FIR Luminosities Corresponding to the Line Detection Limits}
\label{subsec:3.2}

In the high-redshift universe, the luminosity functions of FIR fine-structure lines remain significantly less constrained than those of the total infrared luminosity. 
Consequently, existing luminosity function studies alone are insufficient for estimating the number of galaxies detectable with ATT12.

To overcome this limitation, we convert the line-luminosity detection limits of section~\ref{subsec:3.1} into corresponding infrared luminosities using the empirical relations between line luminosity ($L_{\rm line}$) and infrared luminosity ($L_{\rm IR}$) derived by \citet{Bonato2019}. 
From \textit{ISO} and \textit{Herschel} measurements, they report 
$\log_{10}(L_{\rm line}/L_{\rm IR}) = -2.78$, $-2.92$, and $-2.84$ 
for the \cii~158~$\mu$m, \oiii~88~$\mu$m, and \oiii~52~$\mu$m lines, respectively.
We adopt these values to estimate the infrared luminosity ranges accessible to ATT12, noting that uncertainties in the empirical relations are omitted here because our goal is to obtain first-order sensitivity estimates.

\begin{figure*}[!t]
    \centering
    \includegraphics[width=0.45\linewidth]{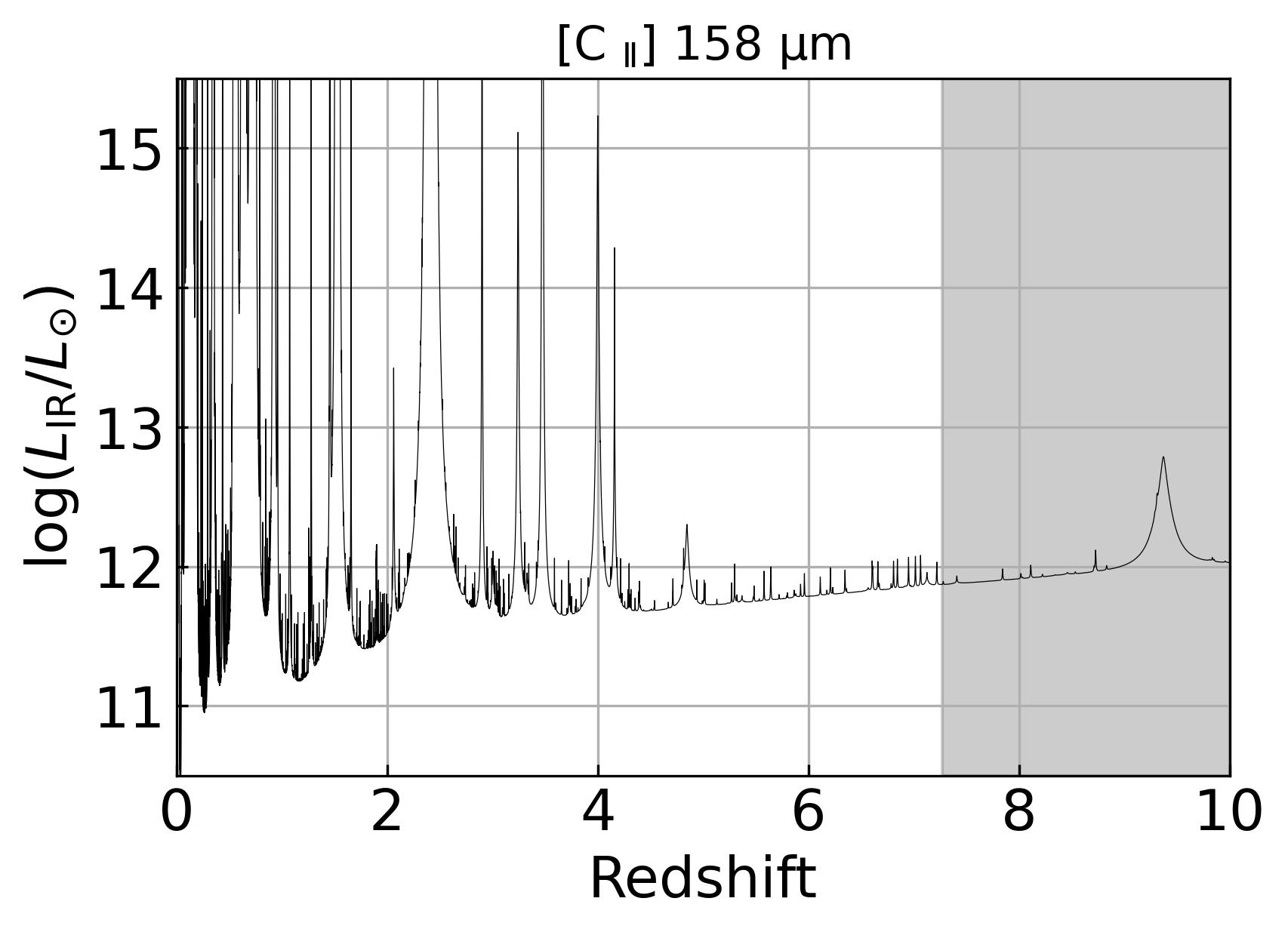}
    \includegraphics[width=0.45\linewidth]{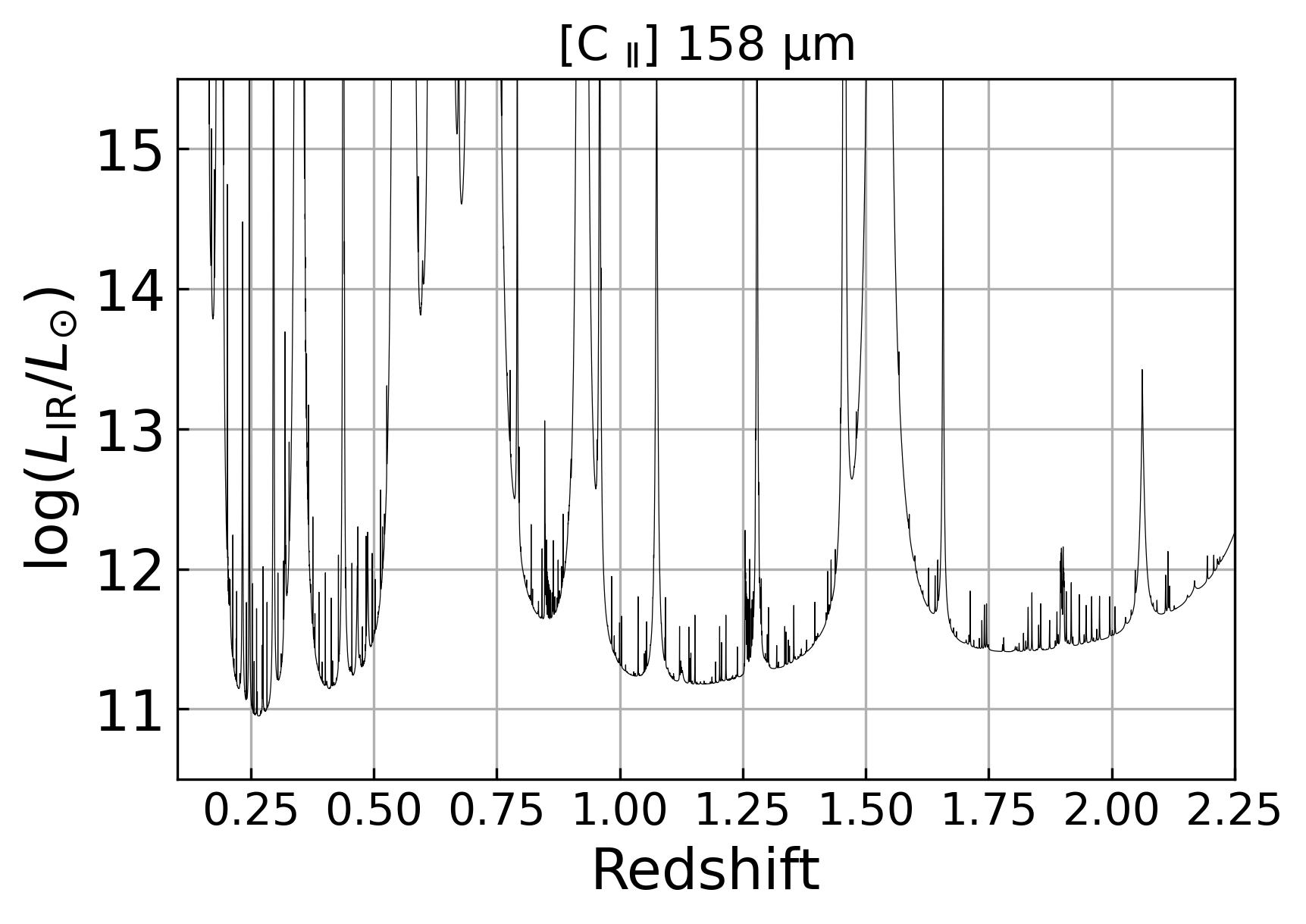}
    \includegraphics[width=0.45\linewidth]{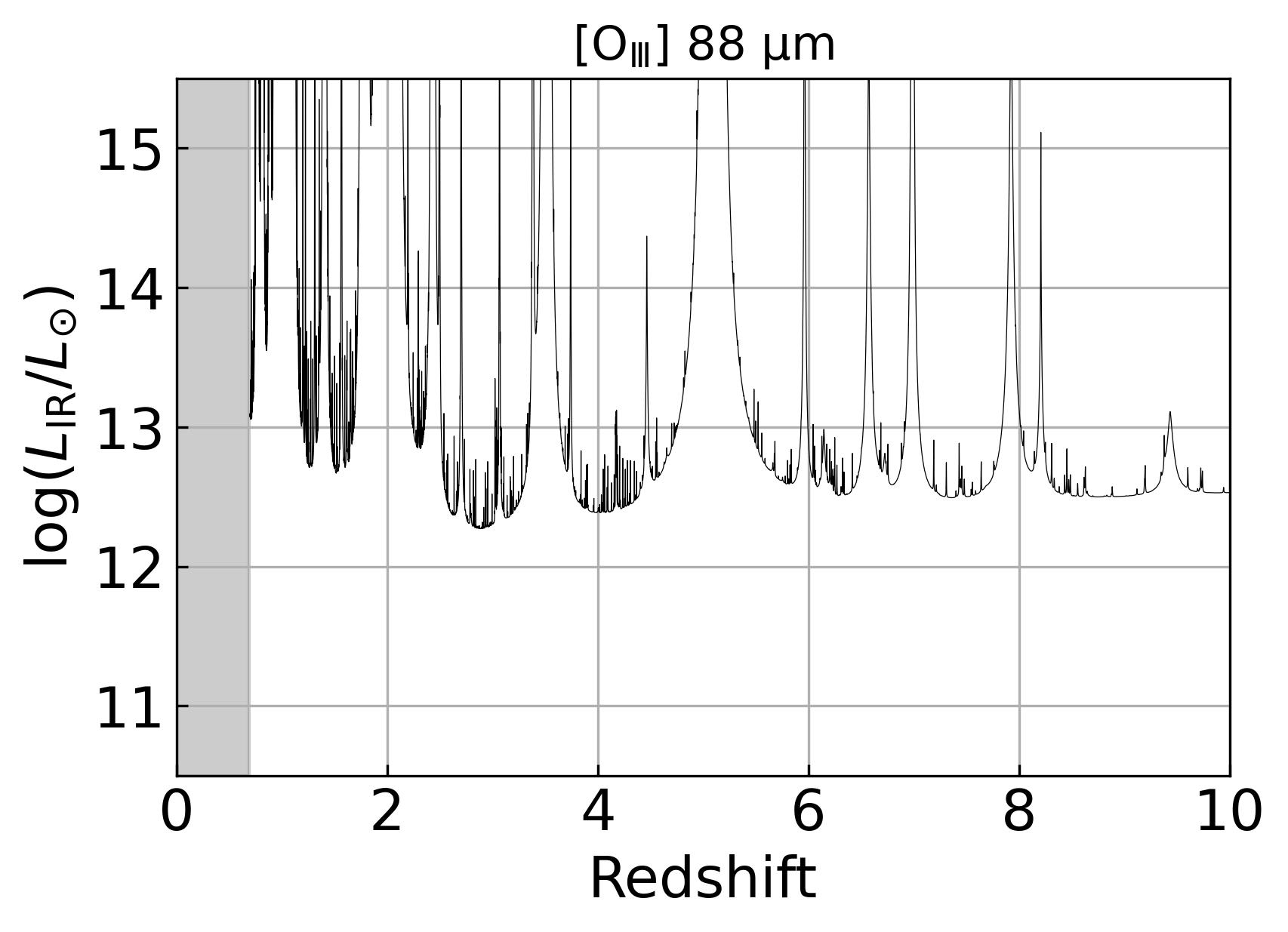}
    \includegraphics[width=0.45\linewidth]{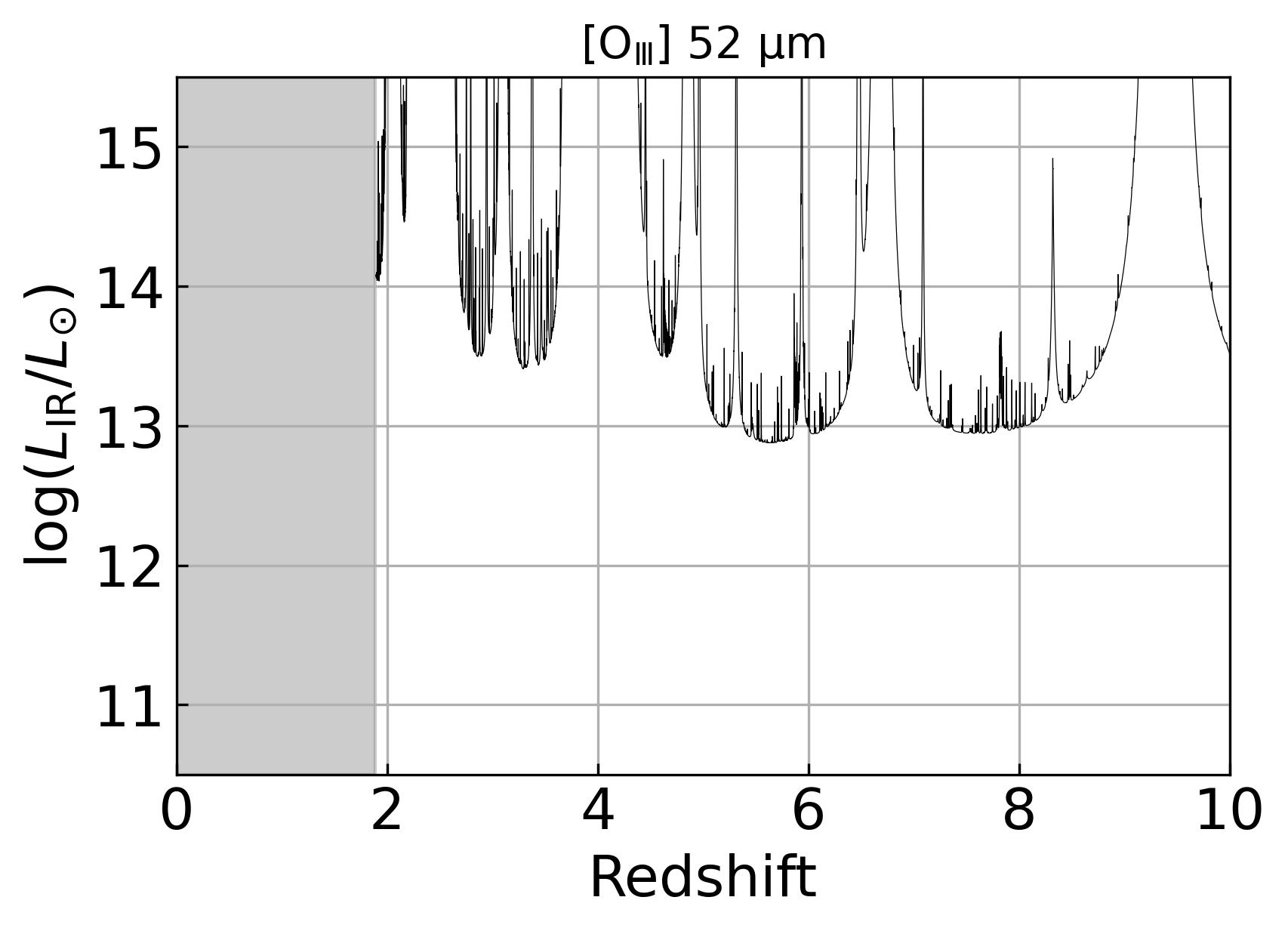}
    \caption{\raggedright
    Infrared luminosity detection limits for 
    [C\,\textsc{ii}]\,158~$\mu$m (top panels), 
    [O\,\textsc{iii}]\,88~$\mu$m (bottom left), 
    and [O\,\textsc{iii}]\,52~$\mu$m (bottom right). 
    Black curves show the 5$\sigma$ limits for 100\,hr observations under the ``winter 50\%'' condition (PWV\,=\,150\,$\mu$m). 
    Shaded regions mark frequencies outside ATT12’s operational range.
    }
    \label{fig:LIR_limit}
\end{figure*}

Figure~\ref{fig:LIR_limit} presents the corresponding detectable infrared luminosities. 
The gray shaded regions denote redshifts (frequencies) inaccessible to ATT12. 
The main results are as follows:

\begin{itemize}
    \item \textbf{[C\,\textsc{ii}]\,158~$\mu$m:}
    ATT12 can detect galaxies with 
    $L_{\rm IR} \gtrsim 10^{11.5} L_\odot$ 
    at $z\simeq0.2$--0.5, 1.0--1.3, and 1.7--2.0.  
    At $z\sim3$--7, sources with 
    $L_{\rm IR} \gtrsim 10^{12} L_\odot$ 
    remain detectable.  
    Thus, ATT12 can reach U/LIRG luminosities from the local universe to the epoch of reionization.

    \item \textbf{[O\,\textsc{iii}]\,88~$\mu$m and [O\,\textsc{iii}]\,52~$\mu$m:}
    The \oiii~88~$\mu$m line is detectable for 
    $L_{\rm IR} \gtrsim 10^{12.5} L_\odot$ 
    at $z\simeq2.5$--3.0, 3.8--4.8, and 6--10.  
    For \oiii~52~$\mu$m, ATT12 detects 
    $L_{\rm IR} \gtrsim 10^{13} L_\odot$ 
    at $z\simeq5.0$--6.5 and 7.0--8.2.  
    These sensitivities enable detections of HyLIRGs during cosmic noon and the reionization era.
\end{itemize}

These results should be regarded as first-order estimates. 
ULIRGs and HyLIRGs exhibit the well-known FIR ``line deficits'' 
\citep[e.g.,][]{Malhotra2001, Gracia-Carpio2011, Diaz-Santos2017, Herrera-Camus2018}, 
in which line-to-IR luminosity ratios decrease with increasing radiation field intensity or compactness. 
Their physical origin remains debated \citep[e.g.,][]{Rybak2019}.  
Furthermore, wide-area surveys reveal that a subset of IR-luminous galaxies are gravitationally lensed, 
in which case large apparent luminosities do not imply intrinsically weak FIR line emission.

Finally, we note that Appendix~\ref{sec:app2} (figure~\ref{fig:LIR_limit_appendix}) presents corresponding infrared-luminosity detection limits for other bright FIR lines, including [N\,\textsc{ii}]\,122~$\mu$m, [N\,\textsc{ii}]\,205~$\mu$m, and [N\,\textsc{iii}]\,57~$\mu$m.

\subsection{Expected Number of Galaxies Detectable with the MKID Camera on ATT12}
\label{subsec:3.3}

In contrast to the previous subsections, which examined the spectroscopic capabilities of ATT12, 
here we turn to the wide-field imaging surveys with the MKID camera. 
As introduced in section~\ref{subsec:intro2}, ATT12 has the capability to survey nearly the entire southern sky 
($\sim$10,000~deg$^2$), excluding regions along the Galactic plane. 
We estimate the number of galaxies that can be detected in such surveys as a function of redshift and luminosity. 
These imaging surveys will play a crucial role in identifying bright sources suitable for spectroscopic follow-up observations with ATT12, 
thereby providing a well-defined sample of luminous DSFGs for detailed ISM diagnostics.

\subsubsection{Infrared Luminosity Functions}
\label{subsubsec:3.3.1}

To estimate the number of galaxies as a function of redshift and luminosity, we adopt IR luminosity functions from the literature. We consider three commonly used parameterizations: the Schechter function \citep{Schechter1976}, the modified Schechter (or Saunders) function \citep{Saunders1990, Takeuchi2003, LeFloch2005}, and the double power-law (DPL) function \citep[e.g.,][]{Fujimoto2024}.
The functional forms of the adopted luminosity functions are given by:
\begin{align}
\phi(L) &= \phi^* \left( \frac{L}{L^*} \right)^{\alpha} 
          \exp\!\left( - \frac{L}{L^*} \right), \\
\phi(L) &= \phi^* \left( \frac{L}{L^*} \right)^{1 - \alpha} 
          \exp\!\left[ - \frac{1}{2\sigma^2} 
          \log^2\!\left( 1 + \frac{L}{L^*} \right) \right], \\
\phi(L) &= \frac{\phi^*}{(L/L^*)^{\alpha} + (L/L^*)^{\beta}},
\end{align}
respectively, where $\phi(L)$ is the number density of galaxies per unit luminosity, $\phi^*$ is the characteristic number density, $L^*$ is the characteristic luminosity, and $\alpha$, $\beta$, and $\sigma$ are parameters that describe the faint- and bright-end slopes and the width of the transition, respectively.

Although IR luminosity functions are well studied compared with FIR line luminosity functions, there remains a considerable scatter among the published literature IR luminosity functions particularly at $z\gtrsim3-4$. To capture this, we adopt several literature values as summarized in table~\ref{tab:appendix_IRLF} in Appendix~\ref{sec:app3}. 

Figure~\ref{fig:lf_examples} shows examples of the IR luminosity functions at $z \sim 6$--7 adopted in this study. While the three functions exhibit similar behavior at low luminosities, they diverge significantly toward the bright end. This difference is critical, as ATT12 is primarily sensitive to luminous infrared galaxies (ULIRGs and HyLIRGs) with $L_{\mathrm{IR}} \gtrsim 10^{12-13}\,L_\odot$. We emphasize that the precise shape of the bright end of the IR luminosity function represents one of the key scientific questions that ATT12 will be able to address in detail.

\begin{figure}[t]
    \centering
    \includegraphics[width=0.95\linewidth]{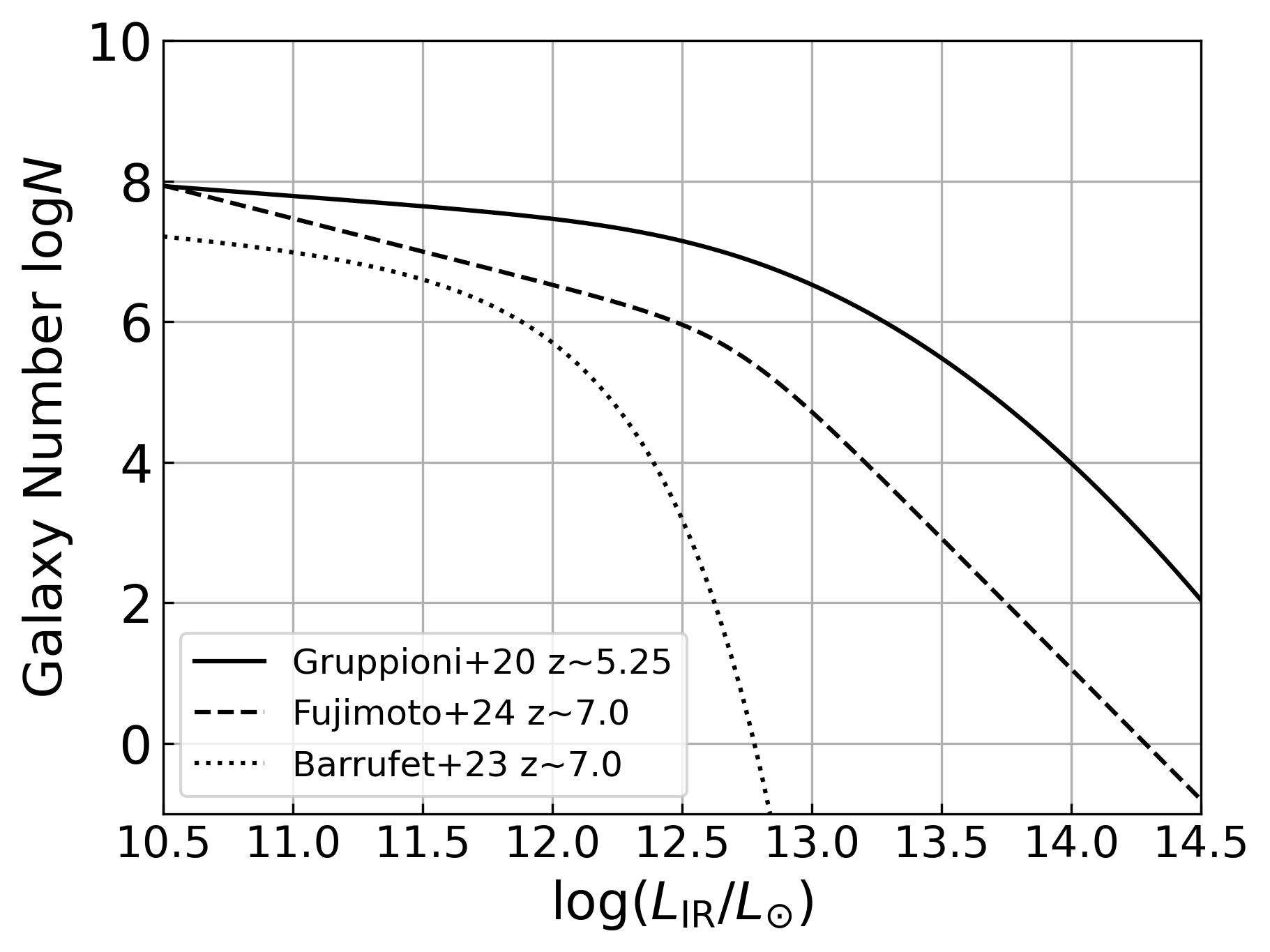}
    \caption{
        \raggedright
        Representative infrared luminosity functions. 
        The solid, dashed, and dotted curves correspond to the parameterizations of 
        \citet{Gruppioni2020}, \citet{Fujimoto2024}, and \citet{Barrufet2023}, respectively. 
        Each curve illustrates one of the redshift bins reported in the corresponding study. 
    }
    \label{fig:lf_examples}
\end{figure}

\subsubsection{Estimate of the Galaxy Numbers Discovered with ATT12}
\label{subsubsec:3.3.2}

For each redshift bin, we first compute the comoving volume of the corresponding spherical shell
from the comoving distance.
By multiplying these comoving volumes by the galaxy number densities derived from luminosity functions, we estimate the number of galaxies that ATT12 can detect at each redshift.

We then evaluate whether such galaxies are observable by considering the confusion limits listed in table~\ref{tab:kids_specs}. 
Following the method of \citet{Totani2002}, we model the observed flux densities of galaxies with a given infrared luminosity ($L_{\rm IR} = 10^{12}\,L_{\rm \odot}$) and dust temperature (42\,K), and compare them with the instrument sensitivity including confusion noise. 
We adopt the $5\sigma$ confusion limits for KIDS-1/2 and the $5\sigma$ sensitivity for 100-hr observations for KIDS-3.   
This approach naturally incorporates the negative $K$-correction in the far-infrared, which keeps the observed flux densities of luminous dusty galaxies nearly constant with increasing redshift, thereby extending the detectable redshift range.

As shown in the left panel of figure~\ref{fig:hirashita} (in private communication with H. Hirashita), galaxies with $L_{\rm IR}=10^{12}\,L_{\odot}$ can be robustly detected by KIDS-1/2/3 up to $z \sim 5$, where the higher-frequency bands are particularly useful for constraining the peak of the dust SEDs.

We also model the observed flux densities of galaxies with a higher infrared luminosity ($L_{\rm IR} = 10^{13}\,L_{\rm \odot}$) and a slightly warmer dust temperature (53\,K), motivated by the positive correlation between $L_{\rm IR}$ and $T_{\rm d}$ (e.g., \citealt{Kovacs2006, Chanial2007, Hwang2010}). As shown in the right panel of figure~\ref{fig:hirashita}, galaxies with $L_{\rm IR}=10^{13}\,L_{\odot}$ can be well detected by KIDS-1/2/3 up to $z \sim 10$--12.

\begin{figure*}[htbp]
    \centering
    \includegraphics[width=0.95\linewidth]{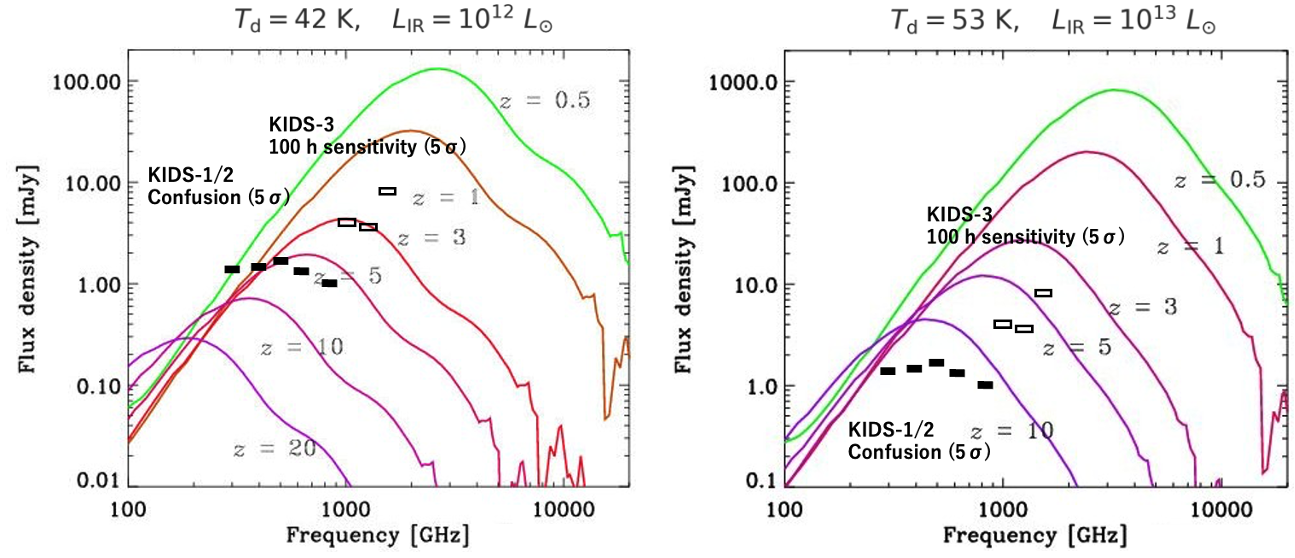}
    \caption{\raggedright
    Dust SEDs with $L_{\rm IR}$ = 10$^{12}$~$L_{\rm \odot}$ and $T_{\rm d}$ = 42 K (left) and $L_{\rm IR}$ = 10$^{13}$~$L_{\rm \odot}$ and $T_{\rm d}$ = 53 K (right), modeled based on \cite{Totani2002} (in private communication with H. Hirashita). In each panel, different color codes indicate different target redshifts. The black filled rectangles indicate the $5\sigma$ confusion limits for KIDS-1/2, whereas open black rectangles correspond to the $5\sigma$ sensitivity at the observation time of 100-hrs. 
          \par}
    \label{fig:hirashita}
\end{figure*}

Figure~\ref{fig:counts} summarizes the predicted number of galaxies detectable with ATT12, based on the infrared luminosity functions compiled in the referenced studies. The corresponding numerical values are provided in table~\ref{tab:The number of galaxies}. 
In the left panel of figure~\ref{fig:counts}, galaxies with $\log(L_{\mathrm{IR}}/L_\odot)=12.0$ are predicted to exceed $10^6$ detectable sources across most of the redshift range $z<5$, although source confusion increasingly limits detections toward higher redshifts. 
The right panel shows that more luminous systems with $\log(L_{\mathrm{IR}}/L_\odot)=13.0$ remain detectable at levels above $10^4$ up to $z\sim7$, beyond which the counts decline rapidly.

These results indicate that ATT12 will enable the construction of large, statistically robust samples of DSFGs, particularly over the key epoch of $z \sim 1$--6. The resulting wide-area census of infrared-luminous galaxies will be highly complementary to ALMA’s high-resolution but narrow-field capabilities.

\begin{figure*}[htbp]
    \centering
    \includegraphics[width=0.45\linewidth]{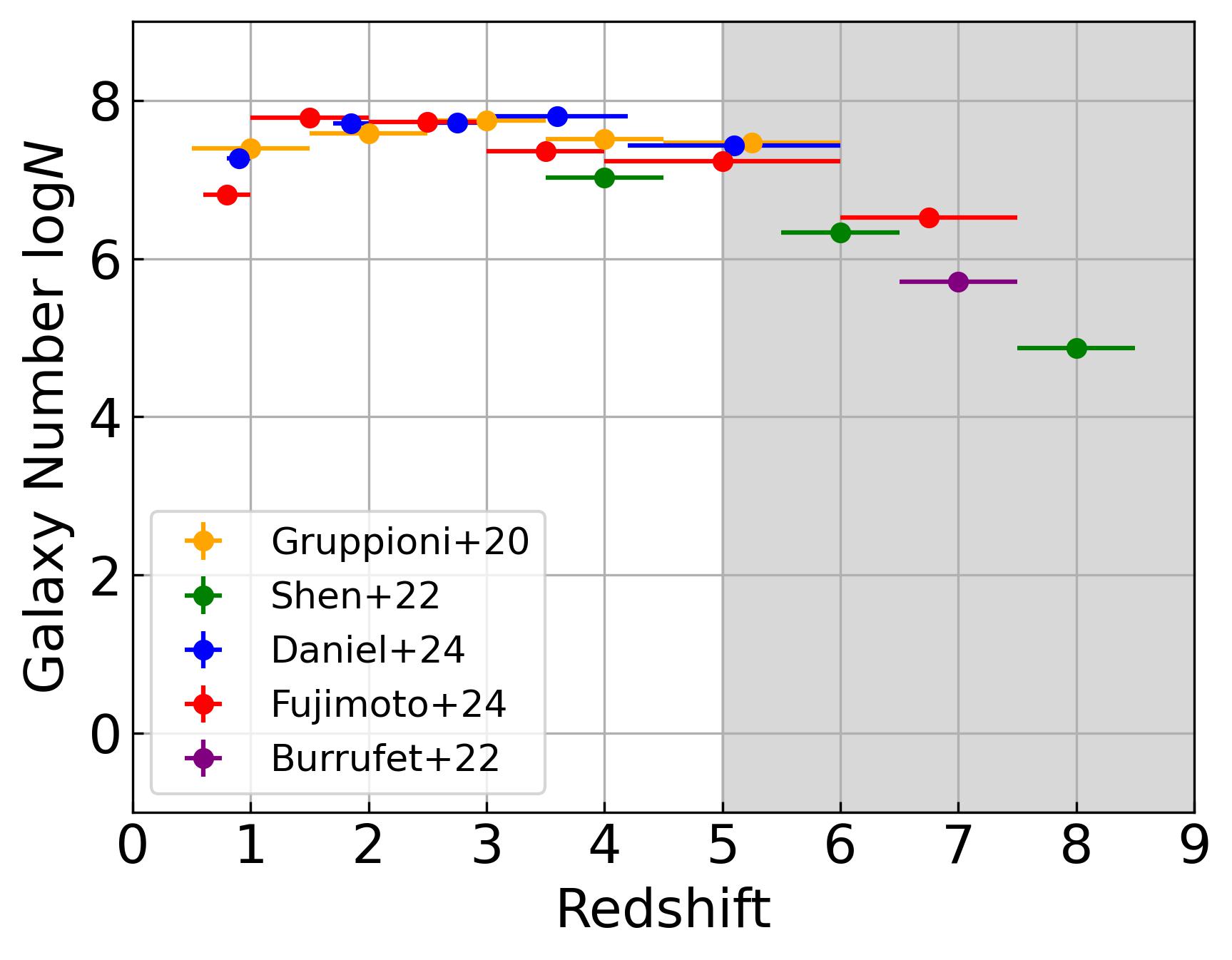}
    \includegraphics[width=0.45\linewidth]{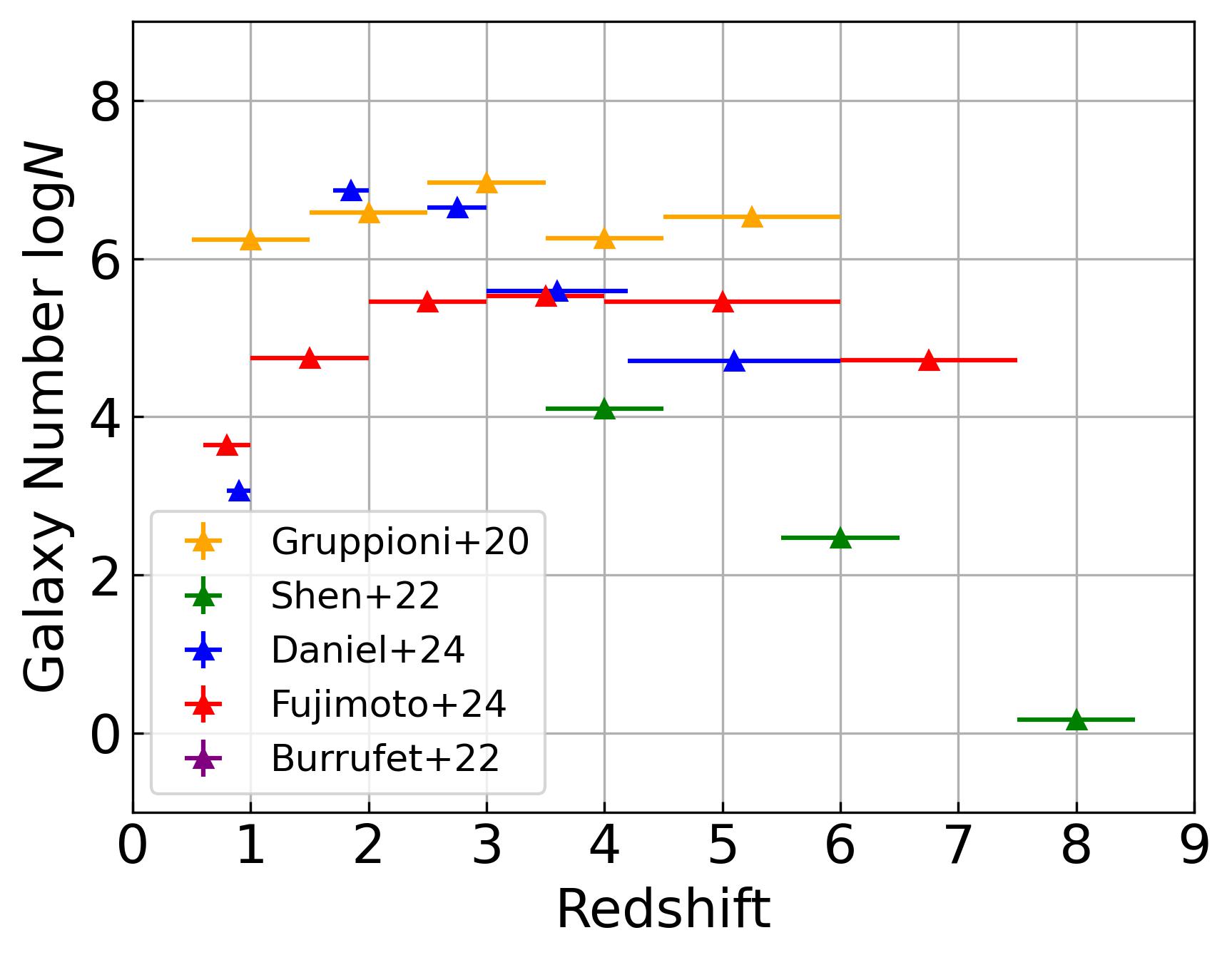}
    \caption{\raggedright
    Number of detectable galaxies in a 10,000~deg$^{2}$ area. 
    The left panel shows galaxies with $\log(L_{\mathrm{IR}}/L_\odot) = 12.0$, 
    and the right panel shows those with $\log(L_{\mathrm{IR}}/L_\odot) = 13.0$. 
    Symbols correspond to models from \citet{Gruppioni2020} (yellow), \citet{Shen2022} (green), 
    \citet{Daniel2024} (blue), \citet{Fujimoto2024} (red), and \citet{Barrufet2023} (purple). 
    Error bars indicate redshift bin widths, and grey shaded areas represent confusion-limited regimes.
    \par}
    \label{fig:counts}
\end{figure*}

\begin{table*}[t]
\tbl{Number of galaxies derived from the infrared luminosity functions.}{%
\centering
\begin{tabular}{lcccccccc}
\hline
Reference & 0.5$\le z<$1.5 & 1.5$\le z<$2.5 & 2.5$\le z<$3.5 & 3.5$\le z<$4.5 & 4.5$\le z<$5.5 & 5.5$\le z<$6.5 & 6.5$\le z<$7.5 & 7.5$\le z<$8.5 \\
\hline\hline
\multicolumn{9}{l}{\textit{$\log(L_{\rm IR}/L_\odot)=12.0$}}\\
\hline
Gruppioni+20 & 7.39 & 7.59 & 7.75 & 7.51 & 7.46 & -- & -- & -- \\
Shen+22      & --   & --   & --   & 7.03 & --   & 6.33 & -- & 4.87 \\
Daniel+24    & 7.27 & 7.71 & 7.72 & 7.80 & 7.43 & -- & -- & -- \\
Fujimoto+24  & 6.81 & 7.78 & 7.73 & 7.37 & 7.23 & -- & 6.52 & -- \\
Barrufet+23  & --   & --   & --   & --   & --   & -- & 5.70 & -- \\
\hline
\multicolumn{9}{l}{\textit{$\log(L_{\rm IR}/L_\odot)=13.0$}}\\
\hline
Gruppioni+20 & 6.24 & 6.59 & 6.96 & 6.26 & 6.53 & -- & -- & -- \\
Shen+22      & --   & --   & --   & 4.10 & --   & 2.47 & -- & 0.17 \\
Daniel+24    & 3.07 & 6.86 & 6.65 & 5.59 & 4.70 & -- & -- & -- \\
Fujimoto+24  & 3.64 & 4.74 & 5.45 & 5.53 & 5.45 & -- & 4.72 & -- \\
Barrufet+23  & --   & --   & --   & --   & --   & -- & 0.00 & -- \\
\hline
\end{tabular}
}
\begin{tabnote}
  \begin{minipage}[t]{0.9\textwidth}
  \raggedright
  {\bf Notes:} The numbers represent the logarithmic number of galaxies per dex (${\rm dex^{-1}}$) derived from the infrared luminosity functions in each reference. 
  Values correspond to the luminosity at $\log(L_{\rm IR}/L_\odot)=12.0$ (upper block) and 13.0 (lower block). 
  For each study, the value at the midpoint of the quoted redshift bin is adopted.
  \end{minipage}
\end{tabnote}
\label{tab:The number of galaxies}
\end{table*}

\section{Discussion}
\label{sec:4}

\subsection{ISM Properties from FIR Diagnostics}
\label{sec:4.1}

Building on the detection limits derived in section~\ref{sec:3}, we now consider the physical information on the ISM that can be obtained with ATT12. In particular, FIR fine-structure line ratios provide powerful diagnostics of electron density, ionization state, and chemical abundances. 

As discussed in section~\ref{subsec:intro2}, ratios of lines with different critical densities, such as \oiii~52/88~\micron\ and \nii~122/205~\micron, are sensitive probes of the electron density \citep[e.g.,][]{ Draine2011, Kewley2019}. Combinations including \oiii\ and \niii\ constrain the oxygen-to-nitrogen abundance ratio (O/N) \citep[e.g.,][]{Pereira-Santaella2017, Kewley2019}. 

Our sensitivity estimates (section~\ref{sec:3}) indicate that such diagnostics become feasible primarily for very luminous systems with $L_{\rm IR} \gtrsim 10^{13}\,L_\odot$, corresponding to HyLIRG-class galaxies. Specifically, \oiii\ line ratios can probe $n_e$ at $z \sim 5$--8, \nii\ ratios at $z \lesssim 5$, and \oiii/\niii\ ratios allow metallicity constraints at $z \sim 4.5$--7 (figures~\ref{fig:LIR_limit} and \ref{fig:LIR_limit_appendix}). The representative conditions are summarized in table~\ref{tab:target_conditions}. 

Although the accessible parameter space is restricted to the most IR-luminous galaxies, these FIR-based diagnostics provide unique insights into the dust-enshrouded ISM. Unlike optical and near-infrared line ratios, which may preferentially trace less obscured regions, FIR lines penetrate deeply into starburst cores \citep[e.g.,][]{Nagao2011, Chartab2022, Spinoglio2022}. By combining FIR diagnostics from ATT12 with optical and near-infrared measurements from JWST, we can obtain a more complete view of ISM conditions from the cosmic noon to the epoch of reionization (e.g., \citealt{Usui2025}).

\begin{table*}[t]
\tbl{Representative ISM diagnostics accessible with ATT12.}{%
\centering
\begin{tabular}{lcc}
\hline
Diagnostic & Accessible $z$ range & Required $L_{\rm IR}$ \\
\hline\hline
\oiii~52/88~\micron\ (electron density) & $5 \lesssim z \lesssim 8$ & $\gtrsim 10^{13}\,L_\odot$ \\
\nii~122/205~\micron\ (electron density) & $z \lesssim 5$ & $\gtrsim 10^{13}\,L_\odot$ \\
\niii~57~\micron\,/\,(\oiii~52~\micron\,$+$\,\oiii~88~\micron) (O/N) & $4.5 \lesssim z \lesssim 7$ & $\gtrsim 10^{13}\,L_\odot$ \\
\hline
\end{tabular}
}
\begin{tabnote}
  \begin{minipage}[t]{0.6\textwidth}
  \raggedright
  The listed redshift ranges are based on the atmospheric windows and sensitivity limits derived in section~\ref{sec:3}. 
  The required luminosities correspond to typical detection thresholds ($L_{\rm IR} \gtrsim 10^{13}\,L_\odot$).
  \end{minipage}
\end{tabnote}
\label{tab:target_conditions}
\end{table*}

\subsection{Statistical constraints on galaxy evolution}
\label{subsec:4.2}

The wide-area survey capability of ATT12 provides a unique opportunity to place statistical constraints on the evolution of DSFGs. 
As shown in section~\ref{sec:3}, galaxies with $\log(L_{\rm IR}/L_\odot)=12.0$ can be detected in numbers exceeding $10^6$ across $z<5$, 
while more luminous systems with $\log(L_{\rm IR}/L_\odot)=13.0$ remain detectable in numbers above $10^4$ up to $z \sim 7$. 
Such large statistical samples are unprecedented in the FIR regime: previous surveys with \textit{Herschel} and SPT probed areas of only a few hundred to a few thousand deg$^2$, 
whereas ATT12 will extend to nearly the full southern sky ($\sim 10^4$ deg$^2$). 
This will allow robust measurements of the bright end of the IR luminosity function, which has so far been poorly constrained at high redshift 
\citep[e.g.,][]{Gruppioni2013,Dudzeviciute2020, Fujimoto2024, Zavala2021}.

Constraining the IR luminosity function at the luminous end is crucial for understanding the relative contribution of dusty galaxies to the cosmic SFRD. 
Recent studies suggest that dusty starbursts play a significant role up to at least $z\sim4$, 
but their importance at earlier times remains uncertain due to limited statistics and selection biases 
\citep[e.g.,][]{Casey2014,RowanRobinson2016, Fujimoto2024, Zavala2024}. 
The large samples accessible with ATT12 will enable a direct comparison between dust-obscured and unobscured SFRDs, 
providing a definitive test of whether dusty star formation contributed significantly already during the epoch of reionization. 
In this way, ATT12 will deliver key insights into the overall balance of obscured and unobscured star formation in shaping galaxy evolution across cosmic time.

\subsection{Complementarity with ALMA, JWST, and PRIMA}
\label{subsec:4.3}

The scientific impact of ATT12 will be maximized through its complementarity with other major facilities. 
ALMA provides exquisite angular resolution and sensitivity, but its narrow field of view makes wide-area surveys impractical. 
ATT12, with its multi-band MKID cameras and large instantaneous field of view, will efficiently identify large samples of DSFGs from the full southern sky ($\sim 10^4$ deg$^2$). 
These samples will form the basis for targeted ALMA follow-up studies of gas kinematics, resolved ISM structure, and molecular gas reservoirs \citep[e.g.,][]{Hodge2020}. 
Thus, ATT12 and ALMA together will enable a complete discovery–characterization sequence from large-scale surveys to high-resolution studies.

In the rest-frame optical and near-infrared, JWST offers powerful diagnostics of metallicity, ionization, and stellar populations \citep[e.g.,][]{Curti2023,Sanders2023, Isobe2023, Nakajima2023}, 
but these measurements are biased toward relatively unobscured regions. 
FIR lines accessible to ATT12 probe the dust-enshrouded ISM and therefore provide a crucial complement. 
At longer wavelengths, the proposed NASA FIR Probe PRIMA will deliver ultra-sensitive photometry and spectroscopy with broad wavelength coverage, 
but with a coarser angular resolution compared to ATT12. 
In this context, ATT12 will occupy a unique niche by combining wide-area coverage, access to the THz regime, and moderate angular resolution from the ground. 
Together with JWST, ALMA, and PRIMA, ATT12 will establish a comprehensive, multi-wavelength framework to study dusty galaxy formation and evolution across cosmic time.

\subsection{Limitations and future prospects}
\label{subsec:4.4}

Despite its unique strengths, ATT12 will also face several limitations. 
First, the attainable sensitivity for faint galaxies is fundamentally restricted by source confusion. 
This effect becomes particularly severe at $z>5$, where individual detections of LIRG-class systems ($L_{\rm IR}\sim10^{12}L_\odot$) are challenging. 
Strategies such as stacking analyses, lensing-assisted fields, and synergy with higher-resolution facilities like ALMA will therefore be essential to extend ATT12’s reach toward fainter populations. 

Second, uncertainties in the FIR luminosity function at high redshift introduce systematic limitations on the expected source counts. 
Current extrapolations beyond $z\sim6$ are poorly constrained, and the true abundance of dusty starbursts during the epoch of reionization remains unclear \citep[e.g.,][]{Casey2014,RowanRobinson2016,Fujimoto2024}. 
ATT12 surveys will directly address this gap, but robust interpretation will require joint analyses with JWST rest-frame UV/optical samples and continued progress in theoretical modeling. 
Another caveat is the so-called FIR line deficit observed in local ULIRGs and HyLIRGs, where line-to-continuum ratios are suppressed at high infrared luminosities. 
Interpreting ATT12 line detections in this regime will therefore require careful consideration of ISM excitation conditions and comparisons with local templates.

\section{Conclusions}
\label{sec:5}

In this paper, we have presented a first feasibility study of ATT12, focusing on its expected performance for studies of DSFGs. Using the instrumental specifications of the heterodyne spectrometers and the MKID imaging cameras, together with atmospheric transmission models for New Dome Fuji, we have assessed the sensitivity, survey capability, and diagnostic potential of ATT12. Our main conclusions are summarized as follows:

\begin{enumerate}
    \item {\bf Spectroscopic sensitivity.}
    Using realistic atmospheric transmission models and the expected performance of the heterodyne spectrometers, we find that ATT12 will detect \cii~158\,$\mu$m emission from galaxies with $\log(L_{\rm IR}/L_\odot)\gtrsim12$ out to $z\sim7$. The \oiii~88\,$\mu$m line will remain observable for HyLIRG-class systems ($\log L_{\rm IR}/L_\odot\gtrsim13$) up to $z\sim10$. 

    \item {\bf FIR luminosities and line diagnostics.}  
    By combining empirical line--to--IR luminosity ratios with our sensitivity estimates, we showed that ATT12 can access \cii, \oiii, and other key transitions in luminous DSFGs. Line ratios such as \oiii~52/88\,$\mu$m, \nii~122/205\,$\mu$m, and \niii~57\,$\mu$m/(\oiii~52+88\,$\mu$m) will enable measurements of electron density and O/N abundance ratios at $z\sim4$--8, although limited to HyLIRG-class systems.

  \item {\bf Imaging survey capability.}
  Wide-field surveys with the KIDS-1 and KIDS-2 cameras will reach confusion-limited depths of $\sim$1--2~mJy at 350--1000~$\mu$m. 
  The KIDS-1 survey, covering 300--500~GHz, is expected to complete its all-sky mapping ($\sim$10{,}000~deg$^2$) within two years, 
  while the higher-frequency KIDS-2 survey (650--850~GHz) will require approximately eight years to achieve similar coverage. 
  From IR luminosity functions, we estimate that ATT12 will detect more than $10^6$ galaxies with $\log(L_{\rm IR}/L_\odot)=12.0$ at $z<5$, 
  and over $10^4$ galaxies with $\log(L_{\rm IR}/L_\odot)=13.0$ up to $z\sim7$. 
  These samples will provide the first statistically robust census of DSFGs across cosmologically representative volumes.

    \item {\bf Scientific implications.}  
    The combination of wide-area imaging and spectroscopic follow-up will allow ATT12 to constrain the bright end of the IR luminosity function and to quantify the contribution of obscured star formation to the cosmic star formation rate density. FIR-based diagnostics will further reveal the physical conditions of dust-enshrouded ISM regions inaccessible to optical surveys.

    \item {\bf Complementarity.}  
    The strengths of ATT12---wide-area THz surveys and moderate angular resolution---are highly complementary to the high-resolution but narrow-field capabilities of ALMA, the rest-frame optical/near-IR diagnostics of JWST, and the ultra-sensitive FIR spectroscopy of the proposed space mission PRIMA. Together, these facilities will provide a comprehensive, multi-wavelength framework to study galaxy formation and evolution across cosmic time.
\end{enumerate}

In summary, ATT12 will open a new discovery space in the THz regime from Antarctica. Its surveys will deliver cosmologically representative samples of DSFGs up to the epoch of reionization, enabling robust constraints on the obscured star formation history and the physical conditions of the ISM in the early universe.

\appendix

\section{Continuum sensitivity of ATT12}
\label{sec:app1}

The detector noise-equivalent power (NEP) adopted in this work is
motivated by the demonstrated performance of existing MKID detectors
and instruments. Laboratory measurements of MKID arrays have demonstrated
detector NEP values of order $\sim10^{-17}\,\mathrm{W\,Hz^{-1/2}}$
\citep{Calvo2016}. Similar sensitivities are realized or expected in
current MKID-based millimeter and submillimeter cameras, including
NIKA2 \citep{Perotto2020}, MUSCAT \citep{Brien2018}, and TolTEC
\citep{Bryan2018}. Although these instruments primarily operate at
millimeter wavelengths ($\sim150$--$300$ GHz), their demonstrated
detector sensitivities indicate that the NEP range adopted in this
work ($10^{-17}$--$10^{-16}\,\mathrm{W\,Hz^{-1/2}}$) is a realistic
assumption for state-of-the-art MKID technology.

NEP of MKID detector arrays under development
at the University of Tsukuba is expected to be 
$\mathrm{NEP}=1\times10^{-17}$, $5\times10^{-17}$, and $1\times10^{-16}\ \mathrm{W\,Hz^{-1/2}}$ 
at $<850$, $850$, and $\ge 1000$ GHz, respectively. 
Corresponding receiver noise temperature is given by

\begin{equation}
T_{\rm RX} = \frac{\mathrm{NEP}}{\sqrt{2}k_{\rm B}\sqrt{B}}
           = \frac{16.2}{\sqrt{B({\rm GHz})}},
             \frac{81.0}{\sqrt{B({\rm GHz})}},
             \frac{162}{\sqrt{B({\rm GHz})}}~{\rm K},
\label{eq:app1_A1}
\end{equation}

respectively, where $k_{\rm B}$ is the Boltzmann constant and $B$ the frequency bandwidth.
The bandwidth is adopted to be $B=35$ GHz at 300 GHz, $10\%$, $5\%$, and $2.5\%$ of the
central frequencies of $400$--$850$ GHz, $1000$--$1500$ GHz, and $2000$ GHz, respectively,
from the atmospheric transmission for $\mathrm{PWV}=0.14$ mm at Dome A in
\citet{H.Yang2010}, where PWV at New Dome Fuji is expected to be nearly the same as that of
Dome A. Resultant $T_{\rm RX}$ are $2.0$--$2.7$ K, $8.8$ K, and $19$--$23$ K at
$<850$, $850$, and $\ge 1000$ GHz, respectively.

The system noise temperature is

\begin{equation}
T_{\rm sys}
=
\frac{
T_{\rm RX}
+
\eta \eta_{\rm op} T_{\rm atm} (1-e^{-\tau})
+
(1-\eta)T_{\rm amb}
}{
e^{-\tau} \eta \eta_{\rm op}
}
{\rm K},
\label{eq:app1_A2}
\end{equation}

where $\tau$ and $T_{\rm atm}$ are the optical depth and temperature of the atmosphere,
$\eta$ the antenna efficiency, $\eta_{\rm op}$ the optical efficiency of MKID,
and $T_{\rm amb}$ the ambient temperature of the antenna.
Expected $T_{\rm sys}$ at each frequency is calculated adopting
$\eta=0.95$, $\eta_{\rm op}=0.3$ ($<850$ GHz), $0.2$ (850 GHz), and
$0.1$ ($\ge1000$ GHz), $T_{\rm atm}\approx T_{\rm amb}\approx200$ K,
and $e^{-\tau}$ at each frequency for $\mathrm{PWV}=0.14$ mm at Dome A
in \citet{H.Yang2010}.

The noise equivalent flux density is

\begin{equation}
\mathrm{NEFD}
=
\frac{2k_{\rm B}T_{\rm sys}}{A\eta_A\sqrt{B}}
=
0.772
\frac{T_{\rm sys}({\rm K})}{\eta_A\sqrt{B({\rm GHz})}}
\ {\rm mJy~s^{1/2}},
\label{eq:app1_A3}
\end{equation}

where $A (=113.1~{\rm m^2})$ is the physical aperture area of the
12-m antenna and $\eta_A$ the aperture efficiency of the antenna at
each frequency.

The minimum detectable flux density is given by

\begin{equation}
\Delta S (1\sigma)
=
\frac{\mathrm{NEFD}({\rm mJy~s^{1/2}})}{\sqrt{t({\rm s})}}
\ {\rm mJy},
\label{eq:app1_A4}
\end{equation}

where $t$ is the integration time.
Table~2 shows five times $\Delta S$ for $t=1$ hour,
adopting $\eta_A$ for the rms surface error of
$\epsilon = 20~\mu{\rm m}$.

\section{FIR Luminosity Detection Limits for \nii\ and \niii\ Lines}
\label{sec:app2}

Following the same method as in section~\ref{subsec:3.2}, we estimate the infrared
luminosity ranges corresponding to the detection limits of the additional FIR
fine-structure lines considered in this appendix, as shown in figure~\ref{fig:LIR_limit_appendix}.
For these transitions, we adopt the empirical relations between line luminosity 
($L_{\rm line}$) and infrared luminosity ($L_{\rm IR}$) derived by \citet{Bonato2019}
from \textit{ISO} and \textit{Herschel} measurements.
They report
\begin{equation}
\log_{10}(L_{\rm line}/L_{\rm IR}) = -3.49,\,-4.09,\,-3.26
\label{eq:app2_1}
\end{equation}

for the [N\,\textsc{ii}]\,122~$\mu$m, [N\,\textsc{ii}]\,205~$\mu$m, and
[N\,\textsc{iii}]\,57~$\mu$m lines, respectively.


\begin{figure*}[htbp]
    \centering
    \includegraphics[width=0.45\linewidth]{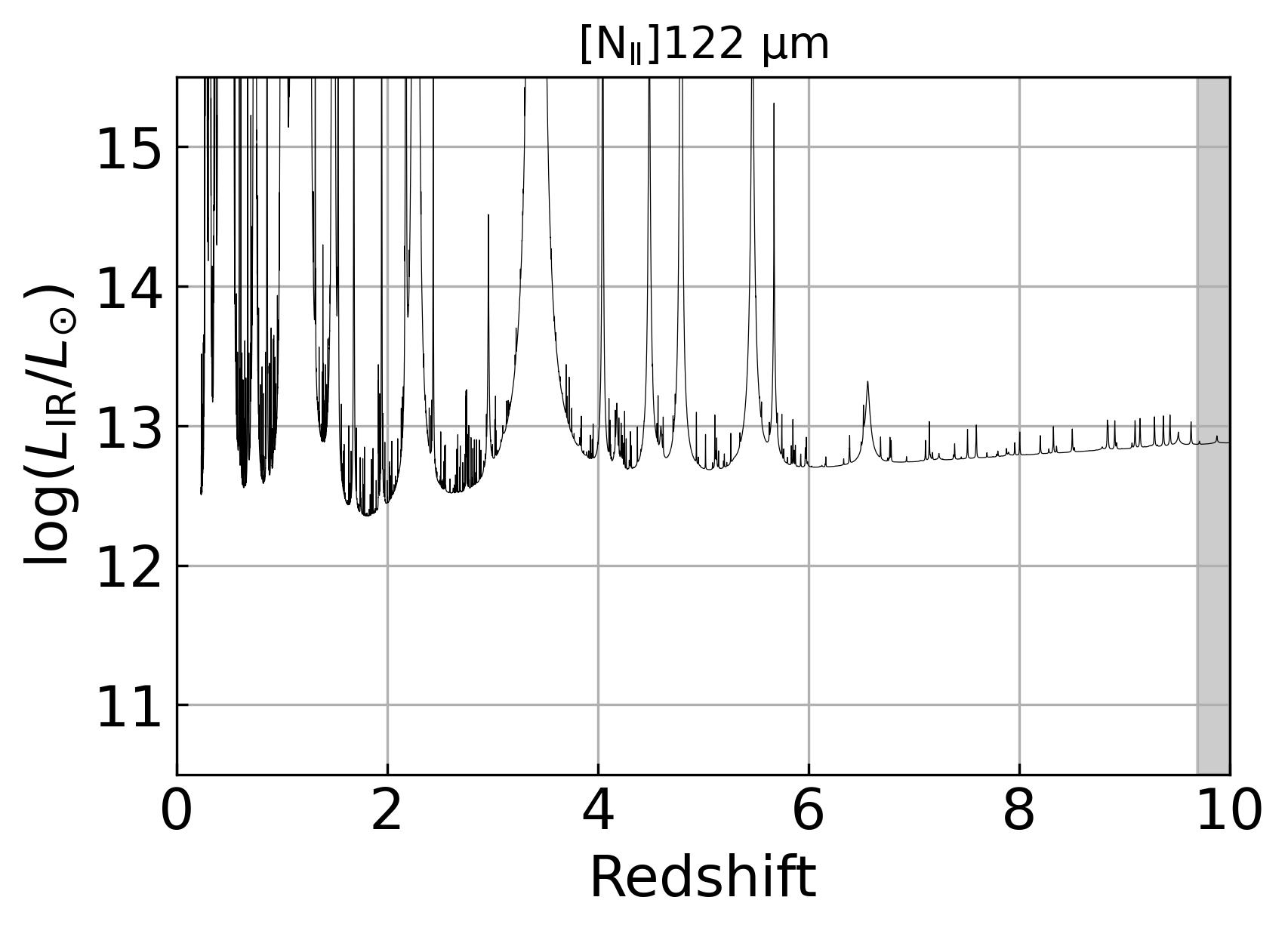}
    \includegraphics[width=0.45\linewidth]{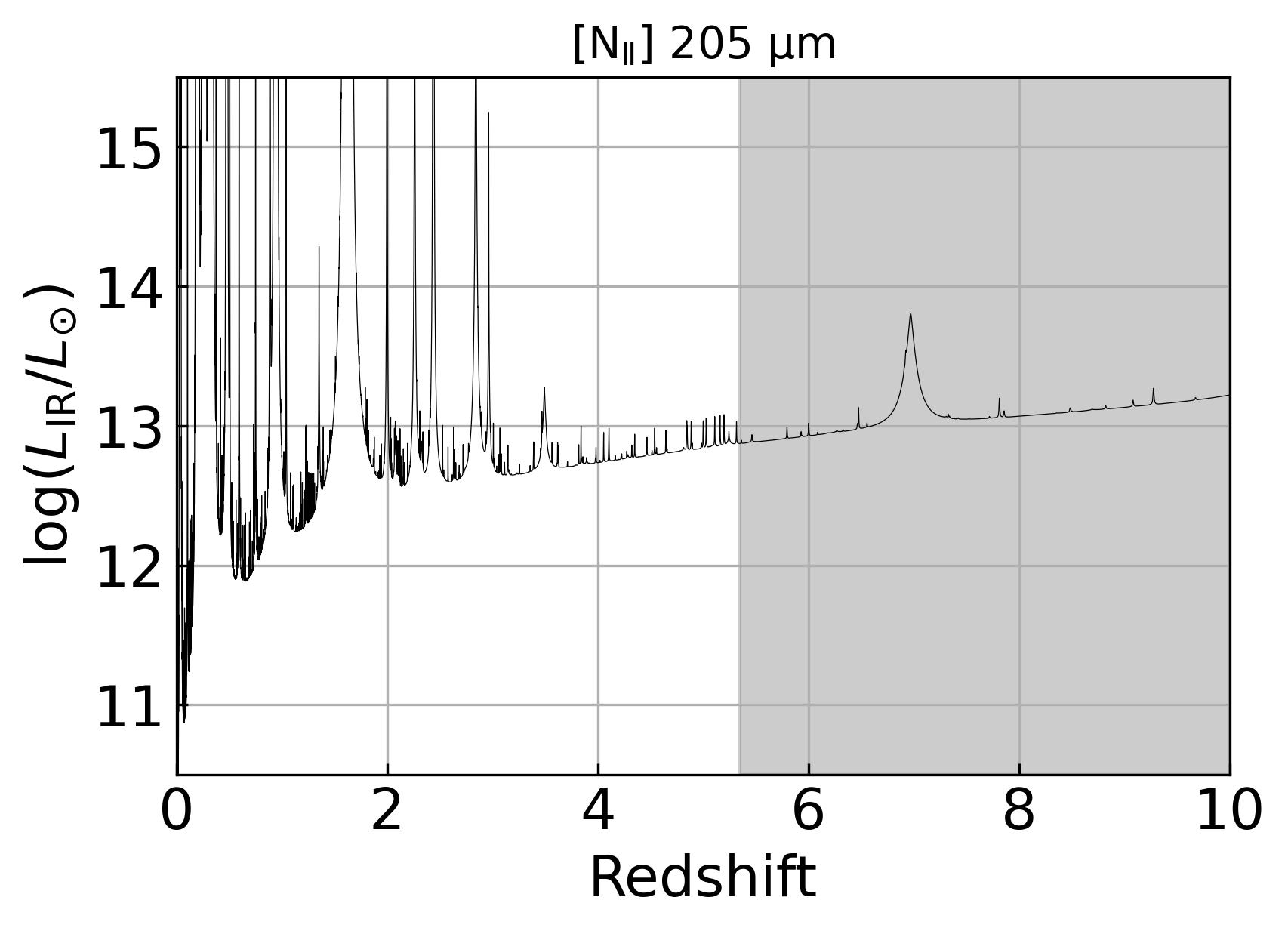}
    \includegraphics[width=0.45\linewidth]{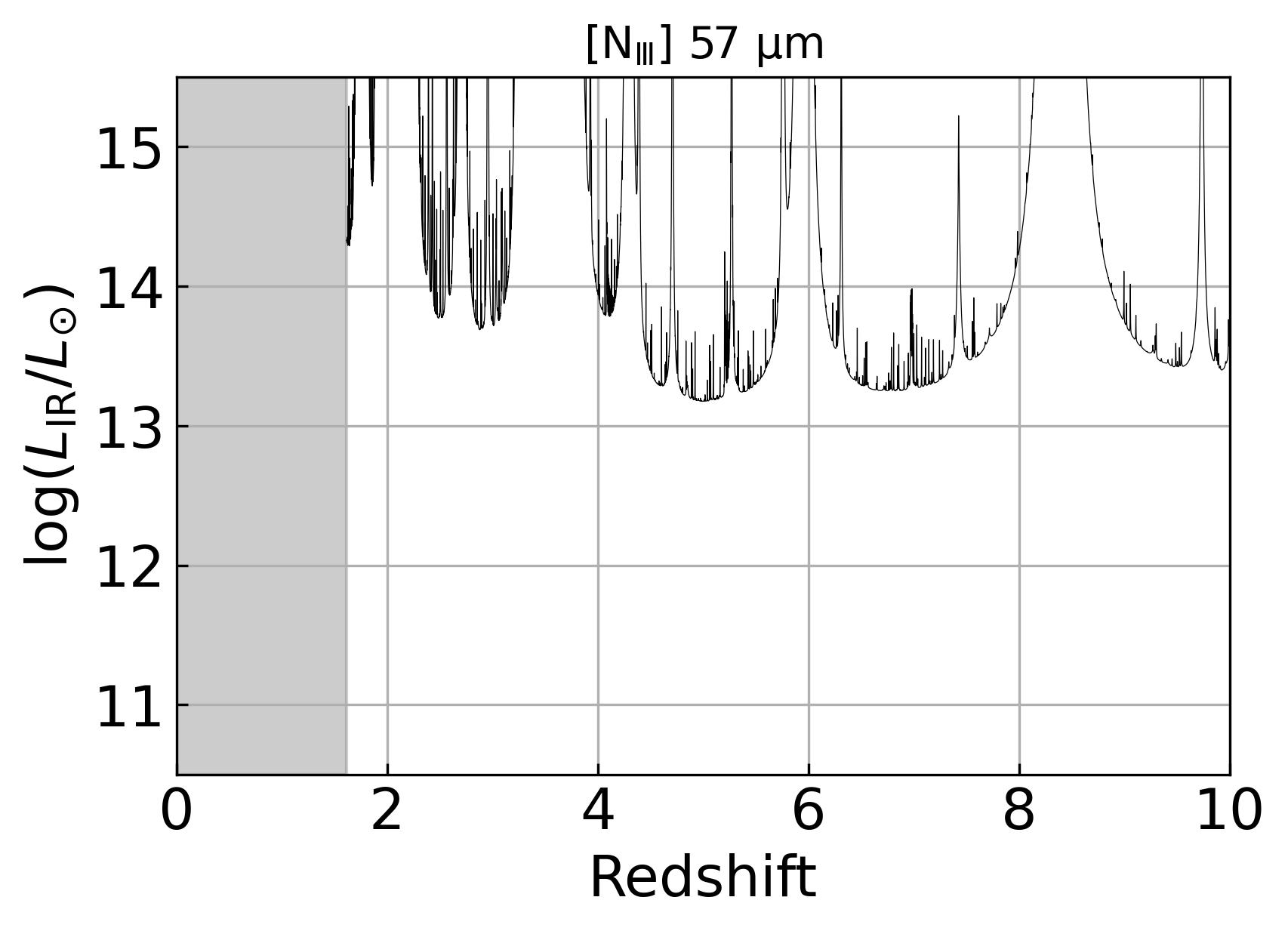}
    \caption{\raggedright
    Infrared luminosity detection limits for 
    [N\,\textsc{ii}]\,122~$\mu$m (top left), 
    [N\,\textsc{ii}]\,205~$\mu$m (top right), 
    and [N\,\textsc{iii}]\,57~$\mu$m (bottom)  
    as a function of redshift.  
    Black curves indicate the 5$\sigma$ detection limits under the ``winter 50\%'' condition 
    (PWV = 150~$\mu$m), and shaded regions mark frequencies outside ATT12’s operational range.
     }
    \label{fig:LIR_limit_appendix}
\end{figure*}

\section{Infrared Luminosity Functions from the Literature}
\label{sec:app3}

We summarize the IR luminosity functions adopted in this study, as shown in table~\ref{tab:appendix_IRLF}. 
For the Schechter function, we use the results of \cite{Barrufet2023}, \cite{Shen2022}, and \cite{Lyon2024}. 

\cite{Barrufet2023} derived the IR luminosity function from 42 galaxies at $z=6.4$--$7.7$, selected via \cii\ 158\,\micron\ line scans (REBELS survey: \citealt{Bouwens2022}). Among these, 16 galaxies were detected in the dust continuum, of which 15 are spectroscopically confirmed through their \cii\ detections.
 
\cite{Shen2022} used the IllustrisTNG simulations (TNG50, TNG100, and TNG300) to investigate infrared luminosity functions of galaxies at high redshift. 
By combining the different volumes and mass resolutions of these simulations, 
they modeled both the bright and faint ends of the luminosity function over $z \gtrsim 4$.

\cite{Lyon2024} decomposed the spectral energy distributions (SEDs) of 22,444 galaxies from the ZFOURGE survey into stellar, dust, and AGN components using the CIGALE code. From the resulting star-forming galaxy sample, they then derived the infrared luminosity function.

For the modified Schechter function, we adopt the results of \cite{Gruppioni2020}, who derived the infrared luminosity function using serendipitous ALMA Band 7 continuum detections from the ALPINE survey. These detections span the ECDFS and COSMOS fields \citep{Le_Fevre2020}, and sample galaxies over a redshift range from $z \sim 0.5$–6.0.

For the double power-law (DPL) luminosity function, we adopt \cite{Fujimoto2024}. 
They analyzed 180 dust continuum sources identified within 33 galaxy cluster fields in the ALMA Lensing Cluster Survey (ALCS), taking advantage of gravitational lensing to probe faint millimeter sources. 
Using the \(1/V_{\mathrm{max}}\) method, they derived IR luminosity functions over the redshift range \(z = 0.6\)–7.5.


\begin{table*}[t]
\centering
\caption{Parameters of infrared luminosity functions from the literature at different redshifts.}
\rotatebox{90}{ 
\resizebox{\textheight}{!}{%
\begin{tabular}{lcccccc}
\hline
Reference & $z$ & $\log_{10}\phi^*$ & $\log_{10}L^*$ & $\alpha$ & $\beta$ & $\sigma$ \\
\hline \hline
Barrufet et al. (2023) & $\sim 7$ 
    & $-4.38$ 
    & $11.60$ 
    & $-1.3$ & -- & -- \\

Shen et al. (2022) & $z \sim 4,6,8$ 
    & $-4.35, -4.70, -5.95$ 
    & $12.25, 12.08, 12.20$ 
    & $-1.73, -1.61, -1.79$ & -- & -- \\

Lyon et al. (2024) & $z \sim 0.9, 1.75, 2.75, 3.6, 5.1$ 
    & $-2.50, -2.96, -3.13, -3.21, -3.61$ 
    & $11.99, 12.78, 12.65, 12.29, 12.19$ 
    & $1.2$ & -- & -- \\
\hline
Barrufet et al. (2023) & $z \sim 1,2,3,4,5$ 
    & $-3.44, -3.45, -3.32, -3.43, -3.73$ 
    & $11.95, 12.01, 12.12, 11.90, 12.16$ 
    & $1.22, 1.15, 1.08, 1.25, 1.28$ 
    & -- & $0.5$ \\

Fujimoto et al. (2024) & $z \sim 0.8, 1.5, 2.5, 3.5, 5.0, 6.75$ 
    & $-3.65, -3.38, -3.62, -4.06, -4.77, -5.26$ 
    & $12.15, 12.52, 12.32, 12.72, 12.73, 12.71$ 
    & $0.94, 0.94, 0.93, 1.04, 0.94, 0.94$ 
    & $3.72$ & -- \\
\hline
\end{tabular}
}}
\label{tab:appendix_IRLF}
\end{table*}

\begin{ack}

We acknowledge Hiroyuki Hirashita for providing us with the information to create a plot in figure 8.
We are grateful to Masato Hagimoto, Hanae Inami, Akio K. Inoue, Hidehiro Kaneda, Kotaro Kohno, Tohru Nagao, Masami Ouchi and Yoichi Tamura for helpful discussions. 
T.H. was supported by JSPS KAKENHI Grant Numbers 23K22529, 25K00020, and 25H00661. T.H. also appreciates support from NAOJ ALMA Scientific Research Grant Number 2025-28A.
 K.M. acknowledges financial support from the Japan Society for the Promotion of Science (JSPS) through KAKENHI grant No. 20K14516. K.M. and T.H. were  supported by JSPS KAKENHI Grant Number 22H01258. 
N.K. was supported by the Grant-in-Aid for JSPS Fellows (25H00661).
H.Y. was supported by MEXT/JSPS KAKENHI Grant Number 21H04489 and JST FOREST Program, Grant Number JP-MJFR202Z. 
M.I. was supported by the Grant-in-Aid for JSPS Fellows (25KJ0670).
Y.N. was supported by MEXT/JSPS KAKENHI Grant Number 23K13140.
\end{ack}

\bibliographystyle{apj}

\end{document}